\documentclass[12pt, oneside, a4paper]{article}
\usepackage{enumitem}
\usepackage{amsmath}
\usepackage{amssymb}
\usepackage{amsthm}
\usepackage{bbm}
\usepackage{amsfonts}
\usepackage{mathtools}
\usepackage{tikz-cd}
\usetikzlibrary{calc}
\usetikzlibrary{shapes,arrows}
\mathtoolsset{showonlyrefs}

\usepackage{etoolbox}

\usepackage[nottoc,notlot,notlof]{tocbibind}

\usepackage{listings}
\usepackage{color}
\definecolor{mygreen}{RGB}{28,172,0}
\definecolor{mylilas}{RGB}{170,55,241}

\newtheorem*{theorem*}{Theorem}
\newtheorem*{remark*}{Remark}

\usepackage{amsthm}
\theoremstyle{remark}

\usepackage{xpatch}

\makeatletter
\xpatchcmd{\@thm}{\thm@headpunct{.}}{\thm@headpunct{}}{}{}
\makeatother

\usepackage{hyperref}
\hypersetup{breaklinks=true}
\usepackage{xurl}
\usepackage[british]{babel}
\usepackage{caption}
\usepackage{subcaption}
\usepackage{graphicx}
\usepackage[justification=centering]{caption}
\usepackage{stackengine}
\usepackage{float}
\usepackage{longtable}
\usepackage{tabularx}
\newcolumntype{Y}{>{\centering\arraybackslash}X}
\usepackage{mathtools}
\usepackage{enumitem}
\usepackage{titlesec}
\titleformat{\chapter}[display]
{\normalfont\huge\bfseries}{}{0pt}{\Huge}
\titlespacing*{\chapter} {0pt}{20pt}{40pt}
\usepackage[thinlines]{easytable}
\usepackage{array}

\usepackage[ruled,vlined,linesnumbered]{algorithm2e}
\usepackage{amsmath}
\usepackage{esvect}

\usepackage[toc]{appendix}

\usepackage{authblk}

\title{\vspace{-1.cm}Synthetic Diagnostic Modeling for Plasma Tomography: Geometry Matrix Computation Methods and Impact of Model Accuracy\\\vspace{-0.2cm}}
\author[1]{D. Hamm\thanks{Corresponding authors: daniele.hamm@epfl.ch, gabriele.partesotti@epfl.ch}}
\author[1]{G. Partesotti${}^{\,*}$}
\author[1]{C. Theiler}
\author[1]{U. Sheikh}
\affil[1]{EPFL-SPC, CH-1015 Lausanne, Switzerland}
\date{}                    
\begin{document}

\vspace{-16.5cm}

\maketitle

\vspace{-0.75cm}
\begin{abstract}
Tomographic emissivity reconstruction from plasma diagnostics data relies on a synthetic model mapping the plasma emissivity to the measured signals. The model, referred to as a geometry matrix in the plasma imaging community, is often built using the line-of-sight (LoS) approximation. This approximation neglects the finite width of the detector viewing beams and can therefore introduce systematic errors. Physically correct volume-of-sight (VoS) models remove this inaccuracy by accounting for the full 3D extent of the viewing beams. Their adoption, however, is sometimes hindered by the difficulty of independently validating them. We present an intuitive and easily inspectable voxel-to-detector (V2D) approach for computing physically accurate VoS geometry matrices, based on discretizing the tokamak vessel into voxels and estimating the contribution of each voxel to the measurements of each detector. We apply the V2D approach to the soft X-ray (SXR) and bolometry systems of the TCV tokamak. Through phantom-based studies on physically realistic emissivity profiles, we quantify the improvement in reconstruction quality obtained by using VoS rather than LoS models. We find that the VoS model yields overall better accuracy and precision; however, interestingly, the simpler LoS model does not introduce a significant systematic bias in the estimated total, core, divertor and main chamber radiated powers. We further compare the V2D geometry matrix with an independent ray-tracing implementation, finding excellent agreement that validates both approaches for routine use at TCV. All routines developed in this work are made openly available.
\end{abstract}
\newpage

\section{Introduction}\label{fwd_mdl_principles}
Fusion devices are often equipped with tomography-capable diagnostics.
This is the case, for example, of systems that measure the radiation from the device over diagnostic-specific photon ranges (soft/hard X-Ray systems, visible light cameras, broadband bolometer detectors).
In first approximation, tomographic data can be interpreted as line-integrated measurements along the lines-of-sight of the diagnostics.
The line-integral interpretation of the measurements of a pinhole camera optical system is rooted in the concept of Radon transform \cite{ingesson_chapter_2008}.
In line-of-sight models, the correct interpretation of the observed tomographic data crucially relies on the notion of etendue \cite{ingesson_chapter_2008}, i.e, the optical throughput of the imaging system.
This quantity accounts for the geometry of the system, describing the amount of light (or radiation) passing through the pinhole and reaching the detector. 
The measurements of a given detector have to be divided by its etendue, in order for them to be interpretable as line-integrated emissivity values.
When working with synthetic emissivity profiles, etendues can be neglected: the employed forward model does not need to be physically exact, as long as the same geometry matrix is used for both generating the synthetic data and performing tomographic reconstruction.
However, when working with experimental data, properly taking into account the geometry of the imaging system is indispensable.
No quantitative imaging is otherwise possible.
\\
As discussed in \cite{ingesson_chapter_2008, jardin_use_2017}, the line-integral approximation with etendue correction is typically expected to provide a rather accurate model for plasma tomographic reconstruction. However, this model remains a simplification of the actual imaging system, strictly accurate only in the ideal limit of infinitely small pinhole and detector. In practice, plasma tomography diagnostics employ pinholes and detectors of finite extent, albeit small. As a consequence, for the case of rectangular pinholes and detectors, the viewing beam, i.e., the spatial sensitivity region $\mathcal{W}\subset \mathbb{R}^3$ of the detector, takes the shape of a truncated rectangular pyramid. The sensitivity function $w:\mathcal{W}\,\to\mathbb{R}^+$, fully characterizing the system, associates to every $\mathbf{x}\in\mathcal{W}$ the fraction of the radiation $\varepsilon(\mathbf{x})$ emitted at $\mathbf{x}$ that reaches the detector. The radiated power $y$ measured by the detector, as we will further discuss in Section \ref{v2d_approach}, is thus given by
\begin{equation}\label{sensitivity_function}
    y = \int_\mathcal{W}w(\mathbf{x})\,\varepsilon(\mathbf{x})\,\mathrm{d}\mathbf{x}\;,
\end{equation}
where the emissivity $\varepsilon(\mathbf{x})$ is expressed in $\mathrm{W/m^3}$. The sensitivity region $\mathcal{W}$ increases with distance from the detector: moving away from the pinhole into the plasma, the beam width increases. The beam width and the overlap between the sensitivity regions of neighboring detectors can be optimized to maximize signal-to-noise ratio and combat aliasing \cite{ingesson_optimization_2002}. In systems developed to estimate 2D emissivity profiles on poloidal cross-sections, such as the RADCAM diagnostic \cite{sheikh_radcam-radiation_2022} installed on the TCV tokamak \cite{Theiler_2026}, the pinholes and detectors often have a larger toroidal than poloidal extent. Since tomographic reconstruction from such systems typically assumes that the emissivity is toroidally symmetric, this design choice promotes accurate signal localization in the poloidal plane while increasing signal-to-noise ratio thanks to the larger toroidal extent of the pinholes and detectors, which increases the amount of radiation that reaches the detectors.\\
The key assumption of the line-of-sight (LoS) approach, as we will further detail in this paper, is that the emissivity is constant over all sections of the viewing beam that are perpendicular to its viewing direction. This direction is given by the beam centerline, which coincides with the sightline of the LoS model. If the beam width is larger than the length scale characterizing the spatial variation of the emissivity function, the LoS approximation loses accuracy. Sharply peaked emissivity profiles, such as those observed for example in the tokamak divertor region, inevitably violate the LoS model assumption. It is therefore important to assess, for a given diagnostic, whether using a LoS model leads to significant systematic errors in the tomographic reconstruction.\\
Physically correct volume-of-sight (VoS) models can be obtained by taking into account the 3D extension of the viewing beams, thus removing the main inaccuracy source of the LoS approximation. Several studies have showed that using VoS models is expected to improve the accuracy of the synthetic diagnostic and the overall tomographic reconstruction quality \cite{ingesson_projection-space_1999, weiland_investigation_2015, carr_description_2018}. Three main approaches have been proposed to account for the finite width of the viewing beams. The first attempts, historically, relied on analytical models operating in projection (sinogram) space \cite{ingesson_optimization_2002, ingesson_projection-space_1999, ingesson_characterization_2000}: while mathematically elegant, 
this strategy cannot handle non-convex vessel geometries \cite{ingesson_characterization_2000} and allows only an approximate treatment of the toroidal beam extension \cite{ingesson_optimization_2002, carr_description_2018}.
The second class of approaches \cite{weiland_investigation_2015, vezinet_non-monotonic_2016} is based on explicitly evaluating the spatially-dependent sensitivity function $w(\mathbf{x})$ at points sampled over the volume of the viewing beam.
The third approach is based on statistical Monte Carlo ray-tracing techniques \cite{carr_description_2018}. In the plasma imaging community, these techniques are often implemented relying on the Python open-source modeling framework Cherab \cite{dr_carine_giroud_2018_1206142}. The ray-tracing approach has become increasingly popular due to its robustness and versatility. It generates a large number of rays from the detector and traces their path through the vessel: the collected statistics are used to compute the geometry matrix (see Section \ref{vos_vs_ray}). The second and third approaches correspond to vessel-to-detector and detector-to-vessel formulations, respectively. Both, if properly implemented, are expected to yield physically accurate models.\\
At TCV, tomographic reconstruction has relied, so far, on a simple LoS model. In recent years, several approaches to implementing VoS models have been explored. However, the absence of a way of validating their physical correctness has prevented their adoption. We here address this issue, discussing and validating VoS models for TCV's RADCAM diagnostic. We recommend the routine adoption of the obtained models, from now on, for tomographic reconstruction from RADCAM's data.\\
The main contributions of this paper are as follows.
\vspace{-0.1cm}
\begin{itemize}
\item We propose an intuitive and easily interpretable vessel-to-detector approach that allows the computation of physically accurate VoS geometry matrices for tomography diagnostics.
\vspace{-0.1cm}
\item We show that our technique can be used to compute VoS models for TCV's SXR and bolometry diagnostics.
\vspace{-0.1cm}
\item We quantify, with phantom-based studies, the expected improvement on tomographic reconstruction quality due to using VoS models for these two diagnostics, rather than simpler LoS ones. We show that the VoS model leads to overall better performance than the LoS one in terms of accuracy and precision of the reconstructed emissivity profile. Interestingly, however, we find that the LoS model does not introduce a significant systematic bias in the estimated total radiated power.
\vspace{-0.1cm}
\item We compare the obtained VoS geometry matrix with an alternative ray-tracing implementation: the excellent agreement between the two approaches validates both for routine use at TCV.
\vspace{-0.1cm}
\item We provide open access to all the routines implemented in the context of this work, accessible at:\\
\url{https://github.com/dhamm97/synthetic_tomo_diags.git}
\end{itemize}
The article is structured as follows. In Section \ref{v2d_approach}, we describe the implementation of our voxel-to-detector (V2D) approach for the computation of the geometry matrix $\mathbf{T}$. The approach is based on discretizing the tokamak vessel into voxels; considering each voxel as a volume emitter, we then estimate the fraction of its emitted radiation that impinges each detector. We discuss the properties of the obtained VoS model, addressing its connection to basic LoS models through the notion of etendue. In Section \ref{los_vs_vos}, we use the V2D approach to compute VoS models for the SXR and bolometry systems installed at TCV, both part of the RADCAM diagnostic. We report the results of studies conducted on physically realistic phantoms, assessing the impact of the accuracy of the employed geometry matrix on tomographic reconstruction. In Section \ref{vos_vs_ray}, we briefly discuss the theoretical foundations of ray-tracing approaches. Then, we compare a Cherab-based implementation of TCV's bolometry geometry matrix to the V2D one computed in Section \ref{los_vs_vos}, showing that the two approaches are equivalent for all practical purposes. In Section \ref{vos_summary}, we summarize and discuss the obtained results.

\section{Voxel-to-detector approach to geometry\\matrix computation}\label{v2d_approach}
\begin{figure}[htbp]
\vspace{-1.5cm}
    \centering
\input{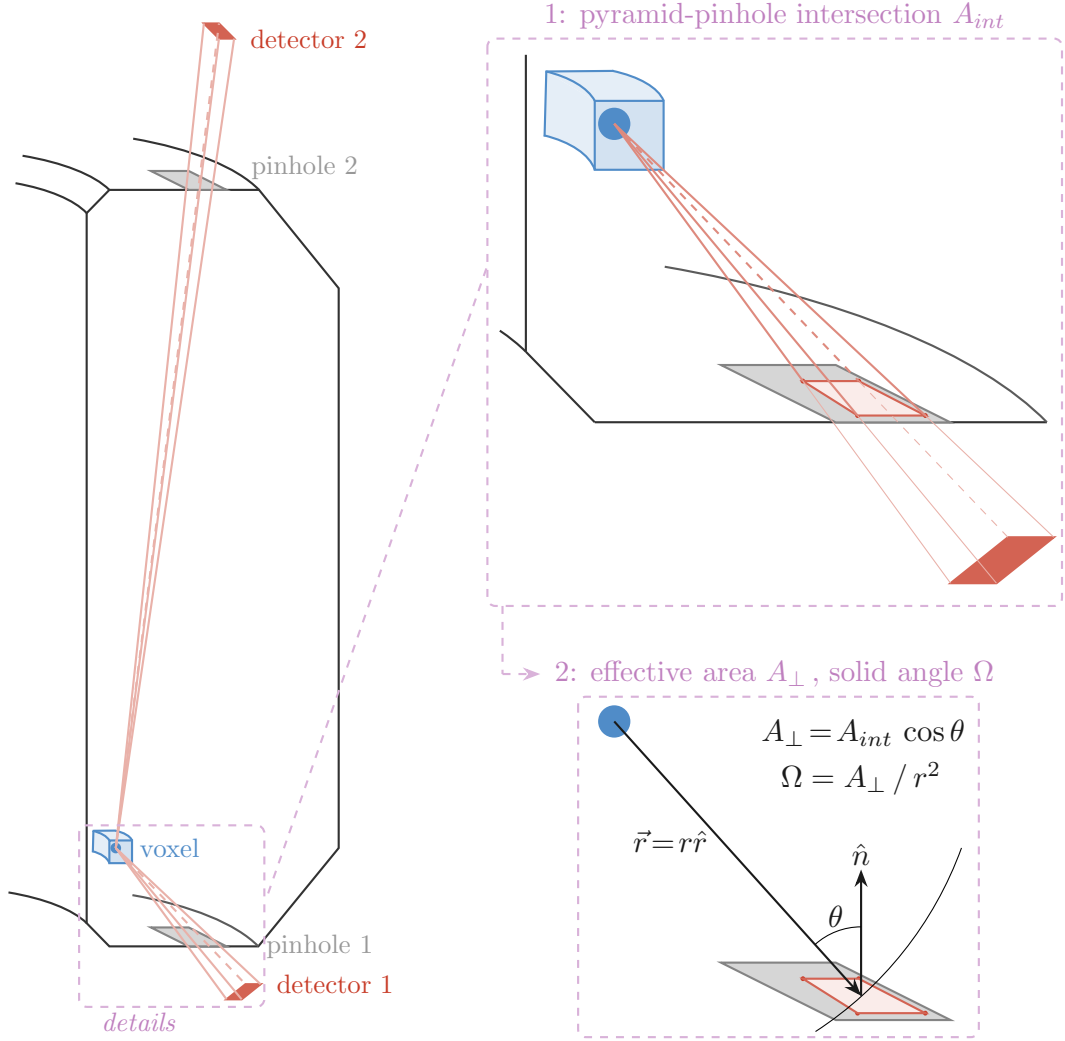}
\vspace{0.5cm}
\captionsetup{justification=raggedright, singlelinecheck=false}
 \caption{Visual representation of the voxel-to-detector (V2D) approach. To the left, a schematic representation of the TCV vessel, a voxel, and two pinhole- detector couples. To the right, details of the bottom part of the figure. Detail 1 illustrates the computation of the intersection area $A_{int}$ between the voxel-detector pyramid and the pinhole. Detail 2 shows the computation of the solid angle $\Omega$ subtended by the detector as seen from the voxel center. This quantity models, for each voxel-detector pair, the fraction of the radiation emitted from the voxel that reaches the detector. The voxel, pinhole and detector sizes are exaggerated for clarity.}
 \label{fig:v2d_approch}
\end{figure}

The vessel-to-detector approach proposed in \cite{weiland_investigation_2015} is based on evaluating the sensitivity function $w(\mathbf{x})$, introduced in Eq.\eqref{sensitivity_function}, on a grid of points belonging to lattice planes parallel to the detector and sectioning the viewing beam. In our voxel-to-detector (V2D) approach, instead, we evaluate $w(\mathbf{x})$ on a discrete grid of 3D voxels. This choice simplifies the definition of the discrete evaluation grid, which is here defined independently of the viewing beam geometry. The evaluation of $w(\mathbf{x})$ is thus done on the same beam-independent voxel grid for all detectors, simplifying the construction of a geometry matrix. We now describe the V2D approach; a visual representation is shown in Fig.\ref{fig:v2d_approch}, while the full algorithm is detailed right after, in Algorithm \ref{alg:v2d}.
Let us consider a voxel $v$ with volume $V_v$, shaped as a 2D poloidal pixel toroidally rotated by an angle $\phi$. If the voxel is sufficiently small that both the sensitivity and emissivity functions can be considered approximately constant over its volume, the contribution $y_v$ of $v$ to the integral in Eq.\eqref{sensitivity_function} is then given by
\begin{equation}\label{y_contrib_formula}
    y_v\approx w(\mathbf{v}) \,\varepsilon(\mathbf{v})\,V_v\;,
\end{equation}
with $\mathbf{v}$ the voxel center. The sensitivity function can be evaluated as follows. Let us consider, for a given rectangular detector $m$, the voxel-detector pyramid $v$-$m$, defined connecting the voxel center to the four corners of $m$. The intersection area $A_{int}$ between the pyramid and the pinhole (see Fig.\ref{fig:v2d_approch}) determines whether $v$ is seen by $m$: if $A_{int}=0$, the voxel is not viewed; if $A_{int}\neq0$, the voxel is viewed and radiation emitted from it contributes to $y$.
\begin{algorithm*}[htbp]
\DontPrintSemicolon
\SetAlgoLined
\SetKwComment{Comment}{\textit{\# }}{}
\SetKwComment{Comment}{\textit{}}{}
\SetCommentSty{textnormal}
\SetKwInOut{Input}{Input}
\SetKwInOut{Output}{Output}
\SetKw{Continue}{continue}
\SetKw{Next}{go to next sub-voxel}
\SetNlSty{}{}{}
 
\caption{Geometry matrix computation: voxel-to-detector (V2D) approach. See Fig. \ref{fig:v2d_approch} for a visual representation.}
\label{alg:v2d}
 
\Input{
  Poloidal grid $(N_z, N_r)$, toroidal resolution $N_\phi$,
  grid refinement factor $h$, $M$ detector-pinhole couples
}
\Output{
  Geometry matrix $\mathbf{T} \in \mathbb{R}^{M \times N},\,N\!=\!N_r N_z\,,\;$ in $[\mathrm{m}^3]$ units
}
 
\BlankLine
\Comment{--- Step 1: Discretization ---}
 
Discretize the poloidal plane into $(N_z, N_r)$ pixels\;
Extrude each pixel toroidally to form $N_r N_z N_\phi$ voxels\;
Subdivide every voxel into $h^3$ sub-voxels ($h$ divisions per side)\;
 
\BlankLine
\Comment{--- Step 2: Geometric factor estimation ---\\Define geometry matrix $\widetilde{\mathbf{F}}\in \mathbb{R}^{M \times (h^3N_zN_rN_\phi)}$ for the refined grid}

\For{each sub-voxel $v=1,\ldots,h^3N_zN_rN_\phi$}{
    \For{each detector $m = 1, \ldots, M$}{
        \Comment{Visibility pre-check to skip non-viewed sub-voxels early}
        Compute intersection between $\vv{v\;\!m}$ line and aperture plane\;
        \lIf{intersection far from pinhole or $\vv{v\;\!m}$ blocked by non-convex structures}{%
            \Next
        }
        Compute intersection $A_{p}$ between $v$-$m$ pyramid and aperture plane; then, area $A_{int}$ of intersection between $A_{p}$ and pinhole\;
        \lIf{$A_\mathrm{int}=0$}{\Next}
        Estimate the solid angle subtended by the effective aperture:
        \vspace{-0.4cm}
        \begin{equation*}
            \Omega_{mv} = A_\mathrm{int} \cos\theta / r^2 
        \end{equation*}\;
        \vspace{-0.9cm}
        Compute the sub-voxel geometric contribution:
        \vspace{-0.4cm}
        \begin{equation*}
            \widetilde{F}_{m v} = (\Omega_{mv} / 4\pi) V_v
        \end{equation*}
        \vspace{-1cm}
    }
} 
\BlankLine
\Comment{--- Step 3: Geometry matrix assembly ---\\Define poloidally refined geometry matrix $\mathbf{F}\in \mathbb{R}^{M \times (h^2N_zN_r)}$, then $\mathbf{T}$}
 
\For{each detector $m = 1, \ldots, M$}{
    \For{each poloidal sub-pixel $i=1,\ldots,h^2N_zN_r$}{
        Sum toroidally over $\mathcal{S}_{tor}^{\,i}=\{\ell_1^{\,i},\ldots,\ell_{hN_{\phi}}^{\,i}\}$,
        $F_{m i} = \sum_{\ell\,\in\,\mathcal{S}_{tor}^{\,i}} \widetilde{F}_{m \ell}$\; 
    }
    \For{each pixel $j=1,\ldots,N_zN_r$ of the reconstruction grid}{
        \Comment{Aggregate sub-pixel contributions}
        Sum poloidally over $\mathcal{S}_{pol}^{\,j}=\{k^{\,j}_1,\ldots,k_{h^2}^{\,j}\}$, $T_{mj}=\sum_{k \,\in\, \mathcal{S}^j_{pol}} F_{mk}$
    }
}
\Return{$\mathbf{T}$, with $T_{mj} \in [\mathrm{m}^3]$}
 
\end{algorithm*}
The amount of this contribution is determined, for voxel $m$, by the solid angle $\Omega_{mv}$ subtended by $A_{int}$ as seen from $v$, which we now compute.
Let $A_\perp=A_{int}\,\cos\theta$ be the effective intersection area, obtained by projecting $A_{int}$ on a plane perpendicular to the $\vec{r}\!=\!r\hat{r}$ line connecting $v$ to the pinhole, with $\theta$ the angle between $\vec{r}$ and the vector $\hat{n}$ normal to the aperture plane (see Fig.\ref{fig:v2d_approch}). The solid angle $\Omega_{mv}$ is thus given by $\Omega_{mv}\approx A_{\perp}/r^2$ (in steradians). Therefore, under the common assumption of isotropic plasma radiation, the unit-less quantity $(\Omega_{mv}/4\pi)$ represents the fraction of the radiation emitted by the voxel that is expected to reach $A_{int}$. In the simplest case of a pinhole that transmits all the incoming radiation, this coincides with the radiation fraction reaching the detector, so that $w(\mathbf{v})=(\Omega_{mv}/4\pi)$ in Eq.\eqref{y_contrib_formula}. The solid angle can also be computed analytically as described in \cite{van_oosterom_solid_1983}, but the approximation $\Omega_{mv}\approx A_{\perp}/r^2$ is extremely accurate for all voxel-detector pairs considered in this work (the resulting error is found to typically be within $0.01\%$).\\
To compute the geometry matrix $\mathbf{T}$, we discretize the poloidal plane into $(N_z, N_r)$ pixels of size $\Delta Z\Delta R$, corresponding to the reconstruction grid to be used for tomographic inversion. For each poloidal pixel $j=1,\ldots,N_zN_r$ with radial location $R_j$, we generate $N_\phi$ voxels by toroidal rotation, as mentioned earlier. The rotation angle $\phi$ is chosen so that $\Delta\phi\approx\Delta R\approx \Delta Z$ with $\Delta\phi=\phi R_j\;$, while $N_\phi$ is chosen large enough to ensure that the sensitivity regions of all $M$ detectors fall within the voxel grid. The resulting grid of $(N_z, N_r, N_\phi)$ voxels must be refined, to ensure that the assumption of constant $w(\mathbf{x})$ and $\varepsilon(\mathbf{x})$ over each voxel volume is sufficiently accurate. Therefore, the grid is refined by a factor $h\in\mathbb{N}$: each voxel is subdivided into $h^3$ sub-voxels, through $h$ divisions in the vertical, radial and toroidal directions. We then define a 3D geometry matrix $\widetilde{\mathbf{F}}\in\mathbb{R}^{M\times(h^3N_zN_rN_\phi)}$ on the refined grid. Its elements $\widetilde{F}_{m\,v}=(\Omega_{mv}/4\pi)V_v$ correspond to the coefficients relating the emissivity of the sub-voxel $v$ to the radiation reaching the detector $m$, as by Eq.\eqref{y_contrib_formula}. With the usual assumption of toroidally symmetric emissivity, a 2D geometry matrix $\mathbf{F}\in\mathbb{R}^{M\times(h^2N_zN_r)}$ is defined on the fine poloidal pixel grid. Its elements are computed, for each poloidal sub-pixel $i=1,\ldots,h^2N_zN_r$, by summing toroidally over all the sub-voxels with indices $S_{tor}^{\,i}=\{\ell^{\,i}_1, \ldots, \ell^{\,i}_{hN_\phi} \}$ whose poloidal projection coincides with the $i$th sub-pixel: $F_{m\,i}=\sum_{\ell\in S_{tor}^{\,i}}\,\widetilde{F}_{m\,\ell}\;$. The matrix $\mathbf{F}$ is a high-fidelity, physically accurate model of the diagnostic. To obtain the coarser geometry matrix $\mathbf{T}\in\mathbb{R}^{M\times NzN_r}$ used for tomographic reconstruction, we aggregate all sub-pixel contributions. Ultimately, the matrix elements for each poloidal pixel $i=1,\ldots,N_zN_r$ are given by $T_{m\,j}=\sum_{k\in S_{pol}^{\,j}}\,F_{m\,k}$, with $S_{pol}^{\,j}=\{k^{\,j}_1, \ldots, k^{\,j}_{h^2} \}$ indexing the sub-pixels of the $j$th poloidal pixel. The matrix $\mathbf{T}$ is the linear forward model defining the tomographic reconstruction inverse problem $\mathbf{y}=\mathbf{T}\boldsymbol{\varepsilon}$, with $\mathbf{y}\in\mathbb{R}^M$ the measurements vector and $\boldsymbol{\varepsilon}\in\mathbb{R}^{N_zN_r}$ the emissivity function discretized on the poloidal pixel grid. The matrix elements are in units of $\mathrm{m^3}$, so that $\mathbf{y}$ is in units of $\mathrm{W}$. In Fig.\ref{fig:sxr_v2d_fine_coarse}, we show the fine and coarse 2D geometry matrices obtained for a subset of the detectors composing RADCAM's SXR diagnostic.\\
\begin{figure}[t]
    \centering
    \begin{tikzpicture}
        \node[anchor=north west, inner sep=0] (img_a) at (0,0) {
            \includegraphics[width=0.185\textwidth]{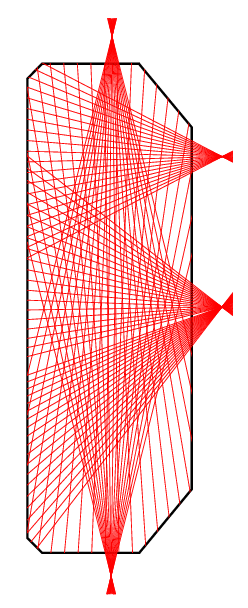}
        };
        \node[anchor=north, font=\small] at (img_a.south) {(a)};

        \node[anchor=north west, inner sep=0] (img_b) at (0.22\textwidth,-3.5mm) {
            \includegraphics[width=0.377\textwidth]{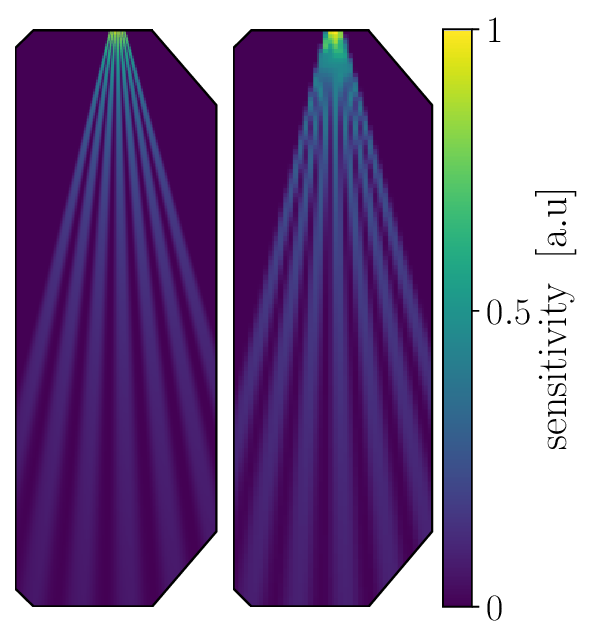}
        };
        \node[anchor=north, font=\small] at ([yshift=-2.3mm]img_b.south) {(b)};
        
        \node[anchor=south, font=\small] at ([xshift=-15mm, yshift=-3.5mm]img_b.north){$\mathbf{F}_{VoS}^{\mathrm{V2D}}$};
        \node[anchor=south, font=\small] at ([xshift=4mm, yshift=-3.5mm]img_b.north){$\mathbf{T}_{VoS}^{\mathrm{V2D}}$};

        \node[anchor=north west, inner sep=0] (img_c) at (0.63\textwidth,-3.5mm) {
            \includegraphics[width=0.377\textwidth]{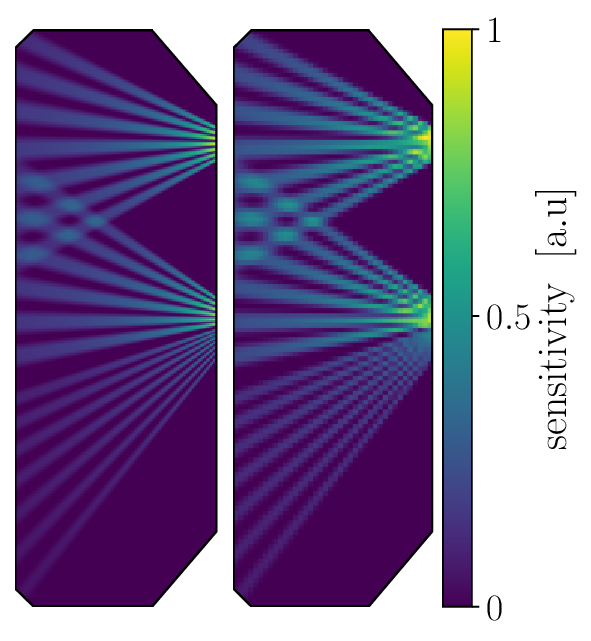}
        };
        \node[anchor=north, font=\small] at ([yshift=-2.3mm]img_c.south) {(c)};
        \node[anchor=south, font=\small] at ([xshift=-15mm, yshift=-3.5mm]img_c.north){$\mathbf{F}_{VoS}^{\mathrm{V2D}}$};
        \node[anchor=south, font=\small] at ([xshift=4mm, yshift=-3.5mm]img_c.north){$\mathbf{T}_{VoS}^{\mathrm{V2D}}$};
    \end{tikzpicture}
    \captionsetup{justification=raggedright, singlelinecheck=false}
    \caption{Voxel-to-detector (V2D) VoS geometry matrix. In Fig.\ref{fig:sxr_v2d_fine_coarse}a, the line-of- sight configuration of RADCAM's SXR diodes. In Figs.\ref{fig:sxr_v2d_fine_coarse}b and \ref{fig:sxr_v2d_fine_coarse}c, visualiza- tions of the fine ($\mathbf{F}_{VoS}^{V2D}$) and coarse ($\mathbf{T}_{VoS}^{V2D}$) versions of the 2D geometry matrix, for a subset of the detectors of the top and lateral cameras, respectively. Arbitrary units are used for visualization purposes.}
    \label{fig:sxr_v2d_fine_coarse}
\end{figure}
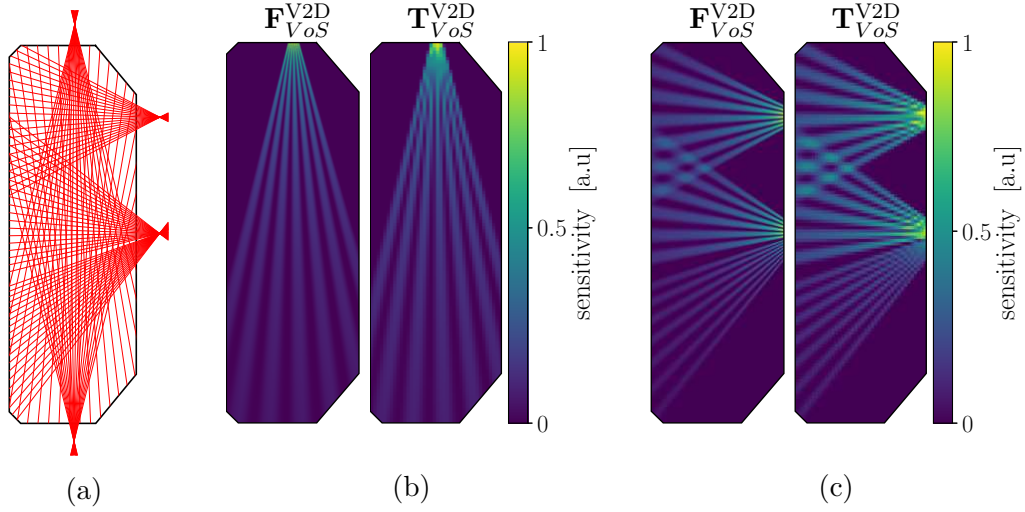
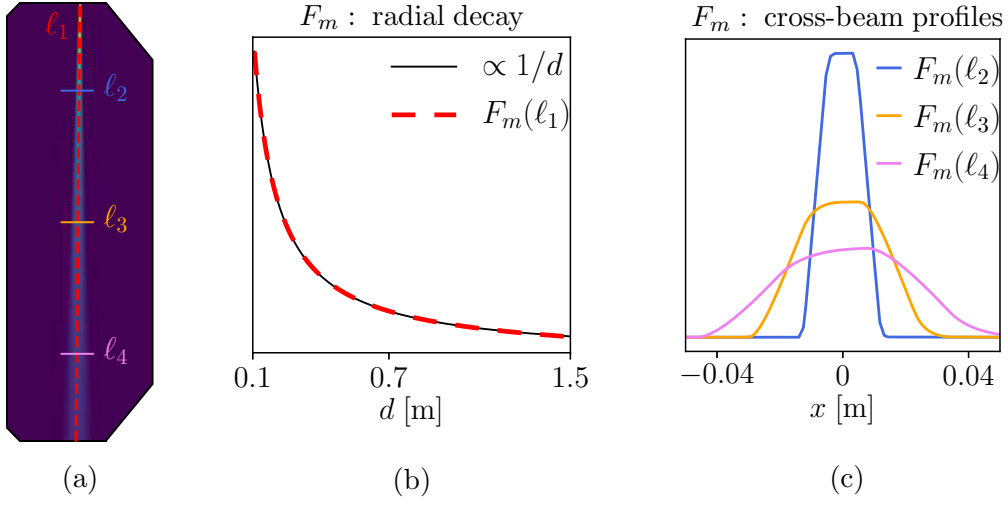
\begin{figure}[htbp]
    \centering
    \begin{tikzpicture}
        \node[anchor=north west, inner sep=0] (img_a) at (0,0) {
            \includegraphics[width=0.17\textwidth]{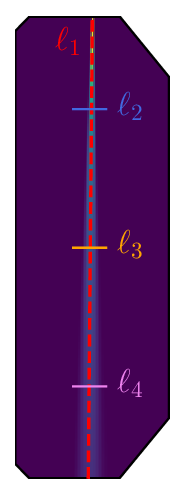}
        };
        \node[anchor=north, font=\small] at (img_a.south) {(a)};

        \node[anchor=north west, inner sep=0] (img_b) at (0.22\textwidth,-5.mm) {
            \includegraphics[width=0.37\textwidth]{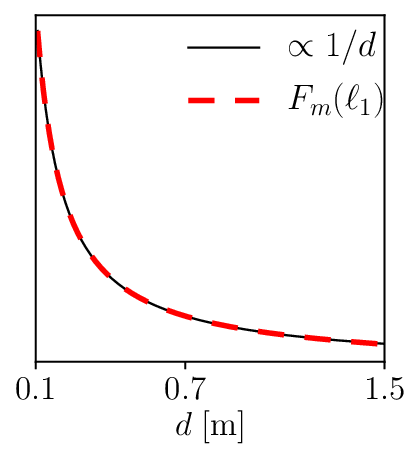}
        };
        \node[anchor=north, font=\small] at ([yshift=-2mm]img_b.south) {(b)};
        
        \node[anchor=south, font=\small] at ([yshift=-2.5mm]img_b.north){$F_m: \; \mathrm{radial\;decay}$};

        \node[anchor=north west, inner sep=0] (img_c) at (0.65\textwidth,-5.25mm) {
            \includegraphics[width=0.35\textwidth]{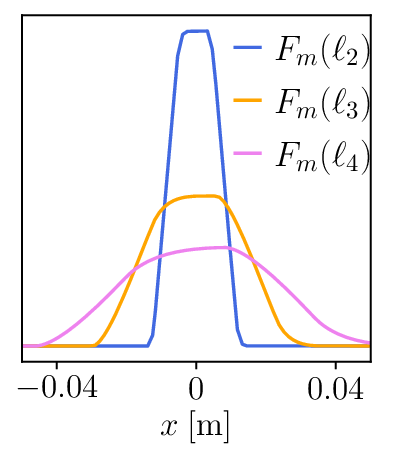}
        };
        \node[anchor=north, font=\small] at ([yshift=-2mm]img_c.south) {(c)};
        \node[anchor=south, font=\small] at ([yshift=-2.5mm]img_c.north){$F_m: \; \mathrm{cross}$-$\mathrm{beam\;profiles}$};
    \end{tikzpicture}
    \captionsetup{justification=raggedright, singlelinecheck=false}
    \caption{Geometry matrix properties. In Fig.\ref{fig:sxr_v2d_eval}a, the geometry matrix $\mathbf{F}$ for detector $m$; overlayed, the evaluation lines $\ell_i$. In Fig.\ref{fig:sxr_v2d_eval}b, the decay rate of $F_m$ as a function of the distance $d$ from the detector. In Fig.\ref{fig:sxr_v2d_eval}c, profiles of $F_m$ along the lines $\ell_2,\ell_3,\ell_4$, as function of the horizontal coordinate $x$.}
    \label{fig:sxr_v2d_eval}
    \vspace{-0.5cm}
\end{figure}
$\!\!\!$To conclude this section, we discuss some properties of the obtained geometry matrix, showing its physical correctness and recovering its connection to LoS formulations. To do so, as shown in Fig. \ref{fig:sxr_v2d_eval}a, we pick a detector $m$ from the top camera of the diagnostic in Fig.\ref{fig:sxr_v2d_fine_coarse}a and evaluate the high-fidelity 2D geometry matrix $\mathbf{F}$ at a large number of points belonging to the lines $\ell_1, \ell_2, \ell_3, \ell_4$. Line $\ell_1$ is the beam centerline, while the three horizontal lines cross the beam at different heights. Bilinear interpolation between pixel values is used to evaluate $\mathbf{F}$ at the sampled points. From Fig.\ref{fig:sxr_v2d_eval}b, we observe that the geometry matrix values decay is perfectly proportional to $1/d$ for this 2D high-fidelity matrix, with $d$ the distance between the evaluation point and the detector $m$. This is the expected theoretical decay rate for an accurate 2D model. Indeed, $1/d$ results from a trade-off between the $1/d^2$ decay rate of the solid angle ($\Omega\propto 1/d^2$) and the rate $d$ at which the toroidal beam width increases with distance from the detector. The latter rate is linear due to the truncated pyramid shape of the viewing beam. In Fig. \ref{fig:sxr_v2d_eval}c, for the three horizontal lines, we plot the geometry matrix values as a function of the distance $x$ between the evaluation point and the beam centerline $\ell_1$; negative values of $x$ indicate points to the left of $\ell_1$. The peakedness of the horizontal profiles, as expected, decreases with distance from the detector, as the beam width increases. However, the integral over $x$ of the profiles is a constant value $I_x$ that does not depend on the chosen line. This property is a direct consequence of etendue conservation \cite{ingesson_projection-space_1999}, which guarantees that the throughput of a lossless optical system is constant along the beam. As the beam width increases, the solid angle subtended by the detector decreases, with the two effects balancing out.
The constant $I_x$ is intimately connected to the etendue of the system $\mathcal{E}$ (expressed in units of $\mathrm{m^2}$), as we now show.
Let $\Pi_d$ be a horizontal plane intersecting the beam at a distance $d$ from the detector, with $\ell_d$ the horizontal line corresponding to the intersection between $\Pi_d$ and the poloidal plane (for non-vertical lines-of-sight, one would consider a plane $\Pi_d$ perpendicular to the beam, with $l_d$ an oblique line).
It then holds
\begin{equation}\label{etendue_conservation}
\begin{aligned}
     \frac{h^2}{\Delta Z\Delta R}I_x &= \int_{\ell_d}\frac{h^2}{\Delta Z\Delta R}F_m(x)\,\mathrm{d}x\\
    &\approx \int_{\Pi_d} \,\frac{\Omega(\mathbf{s})}{4\pi}\,\mathrm{d}\mathbf{s}\\
    &\approx \int_{\Pi_d} \cos\theta(\mathbf{s})\,\frac{\Omega(\mathbf{s})}{4\pi}\,\mathrm{d}\mathbf{s} = \mathcal{E}\;.
\end{aligned}
\end{equation}
The first approximate equality in Eq.\eqref{etendue_conservation} follows from the definition of the elements of the 2D matrix $\mathbf{F}$. Indeed, since the elements of the 3D matrix $\widetilde{\mathbf{F}}$ are given by $\widetilde{F}_{mv}=(\Omega_{m\,v}/4\pi)V_v$ and since $F_m$ is obtained by toroidal summation of the $\widetilde{F}_{mv}$ contributions, $F_m(x)$ normalized by the poloidal sub-pixel size $(\Delta Z \Delta R)/h^2$ represents the toroidal integral of the sensitivity function $w(\mathbf{x})=\Omega(\mathbf{x})/4\pi$. Further integrating this quantity along $\ell_d$ thus yields a numerical approximation of the integral of $w(\mathbf{x})$ over the plane $\Pi_d$, i.e., the second line in Eq.\eqref{etendue_conservation}. The second approximate equality in Eq.\eqref{etendue_conservation} follows because, for relatively narrow beams such as those considered here, it holds $\cos\theta(\mathbf{s})\approx 1 \;\forall\,\mathbf{s}\in\Pi_d\;\mathrm{s.t.}\;w(\mathbf{s})>0$. With the recovered $\cos\theta(\mathbf{s})$ factor, the obtained integral is the general definition of the etendue of the system. In plasma tomography, the etendue is often operatively defined without the $\cos\theta(\mathbf{s})$ factor \cite{jardin_use_2017, ingesson_projection-space_1999}. The fact that $\int_{\ell_d}F_m(x)\mathrm{d}x=I_x$ for all distances $d$ shows that the etendue is numerically conserved by the model, confirming its physical correctness. The coarse resolution matrix $\mathbf{T}$ preserves this property.\\
If the assumption of constant emissivity over all viewing beam sections is satisfied, with $\varepsilon(\mathbf{s})=\varepsilon_d\;\forall\mathbf{s}\in\Pi_d\cap\mathcal{W}$, the VoS and LoS formulations are equivalent from a measurement point of view. Indeed, by defining $\mathcal{E}_{LoS}$ as the second line of Eq.\eqref{etendue_conservation} and writing volume integration $\int_\mathcal{W}\mathrm{d}\mathbf{x}$ as $\int_{d}\int_{\Pi_d}\mathrm{d}\mathbf{s}\,\mathrm{d}s_{_{LoS}}$ (i.e., line-integrating along the LoS the integrals over planes $\Pi_d$), Eq.\eqref{sensitivity_function} can be written as
\begin{equation}\label{los_vos_equivalence}
\begin{aligned}
    y &= \int_{\mathcal{W}}w(\mathbf{x})\varepsilon(\mathbf{x})\,\mathrm{d}\mathbf{x}\\
    &= \int_{d}\varepsilon_d(s_{_{LoS}})\int_{\Pi_d} w(\mathbf{s})\,\mathrm{d}\mathbf{s}\;\mathrm{d}s_{_{LoS}}\\
    &= \mathcal{E}_{LoS} \int_{d}\varepsilon_d(s_{_{LoS}})\;\mathrm{d}s_{_{LoS}}\;.
\end{aligned}
\end{equation}
The expression derived in Eq.\eqref{los_vos_equivalence} underpins the interpretation of tomographic measurements as line-integrated emissivity values. However, as we discussed in Section \ref{fwd_mdl_principles}, the LoS model assumption of constant emissivity is not true, in general. In Section \ref{los_vs_vos}, we investigate whether relying on this simplified model leads to systematic errors in tomographic reconstruction.

\section{Impact of synthetic diagnostic accuracy on tomographic reconstruction}\label{los_vs_vos}
In this section, we assess the importance of using a physically correct forward model for tomographic reconstruction from data from RADCAM's SXR and bolometry systems. To do so, we proceed as follows.\\
First, we compute the VoS geometry matrix $\mathbf{T}_{VoS}^{V2D}$ using the V2D approach described in Section \ref{v2d_approach}. This model, corresponding to the physically correct synthetic diagnostic, is assumed to be fully accurate and used as reference. We also compute the LoS geometry matrix $\mathbf{T}_{LoS}$. 
The LoS matrix is computed directly on the coarse reconstruction grid with the usual pixel-based approach \cite{jardin_use_2017}. The length of the intersections between the sightlines and pixels are computed, then each matrix row $T_{LoS,\,m}$ is multiplied by the estimated etendue $\widehat{\mathcal{E}}_{LoS,\,m}$ of the $m$th detector. The common \cite{carr_description_2018} approximation
\begin{equation}\label{etendue_los}
\widehat{\mathcal{E}}_{LoS}=\frac{A_{pinhole}A_{detector}\cos\alpha\,\cos\beta}{4\pi L^2}\;,
\end{equation}
is used to estimate the etendue, with $L$ the detector-pinhole distance, $\alpha$ the angle between sightline and pinhole normal $\hat{n}_{pinhole}$ and $\beta$ the angle between sightline and detector normal $\hat{n}_{detector}$.\\
We conduct phantom-based studies to investigate the influence of the geometry matrix on the reconstruction. Studies on physically realistic emissivity profiles are a common and valuable validation tool, allowing the obtained results to be compared against a known ground truth, which is not possible when working with experimental data.
We apply the geometry matrices $\mathbf{T}_{VoS}^{V2D}$ and $\mathbf{T}_{LoS}$ to the phantom profiles, obtaining the tomographic measurements $\mathbf{y}_{VoS}$ and $\mathbf{y}_{LoS}$, respectively.
The signed relative error $\delta_y=(y_{LoS}-y_{VoS})/y_{VoS}$ is used to estimate the expected error committed when approximating the true measurements $\mathbf{y}_{VoS}$ with $\mathbf{y}_{LoS}$.
The measurements $\mathbf{y}_{VoS}$ are then corrupted with Gaussian channel-independent noise, and provided as input for two separate tomographic reconstructions: first using the correct matrix $\mathbf{T}_{VoS}^{V2D}$, then using $\mathbf{T}_{LoS}$.
In the second case, there is a mismatch between the physically correct model used to generate the synthetic tomographic data and the one employed for reconstruction.
We then compare the two obtained reconstructions $\mathbf{x}_{_{MAP}}^{VoS}$ and $\mathbf{x}_{_{MAP}}^{LoS}$ to the ground truth phantom profile $\mathbf{x}$. In both cases, we use the reconstruction model in Eq.(24) from \cite{hamm_tomography_2025}, which features a Gaussian likelihood, an anisotropic diffusion smoothing prior with anisotropic parameter $\alpha=10^{-2}$, and a positivity constraint. To assess the quality of the two inversions, we compute the relative errors $\delta_{P_{rad}}=(P_{rad}(\mathbf{x}_{_{MAP}})-P_{rad}^{true})/P_{rad}^{true}$ on the reconstructed radiated power and the MSEs $E_{_{MAP}}=1/N\;\sum_{i=0}^{N-1}((\mathbf{x})_i-(\mathbf{x}_{_{MAP}})_i)^2$. The study is designed to reproduce the mismatch taking place when using a simplified LoS model to reconstruct experimental data. In Section \ref{sxr_effect}, we focus on the SXR diagnostic, while Section \ref{bolo_effect} is dedicated to the bolometry diagnostic.
\subsection{SXR reconstruction}\label{sxr_effect}
\begin{figure}[htbp]
    \centering
    \vspace{1.25cm}
    \begin{tikzpicture}
        \node[anchor=north west, inner sep=0] (img_a) at (0,0) {
            \includegraphics[width=0.185\textwidth]{sxr_LoS_configuration.eps}
        };
        \node[anchor=north, font=\small] at (img_a.south) {(a)};

        \node[anchor=north west, inner sep=0] (img_b) at (0.22\textwidth,0) {
            \includegraphics[width=0.37\textwidth]{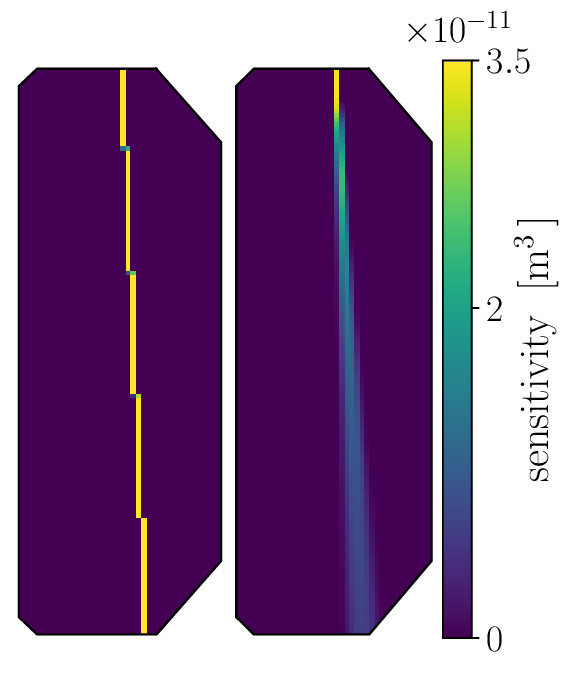}
        };
        \node[anchor=north, font=\small] at ([yshift=-2.9mm]img_b.south) {(b)};
        
        \node[anchor=south, font=\small] at ([xshift=-15mm, yshift=-6.5mm]img_b.north){$\mathbf{T}_{LoS}$};
        \node[anchor=south, font=\small] at ([xshift=4mm, yshift=-6.5mm]img_b.north){$\mathbf{T}_{VoS}^{\mathrm{V2D}}$};

        \node[anchor=north west, inner sep=0] (img_c) at (0.63\textwidth,0) {
            \includegraphics[width=0.37\textwidth]{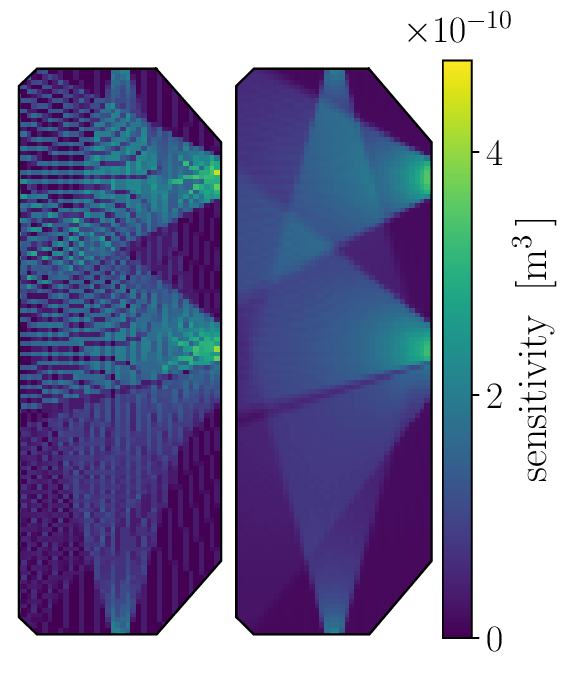}
        };
        \node[anchor=north, font=\small] at ([yshift=-2.9mm]img_c.south) {(c)};
        \node[anchor=south, font=\small] at ([xshift=-15mm, yshift=-6.5mm]img_c.north){$\mathbf{T}_{LoS}$};
        \node[anchor=south, font=\small] at ([xshift=4mm, yshift=-6.5mm]img_c.north){$\mathbf{T}_{VoS}^{\mathrm{V2D}}$};

        \node[anchor=north west, inner sep=0] (img_d) at (0.5,-7.5) {
            \includegraphics[width=0.31\textwidth]{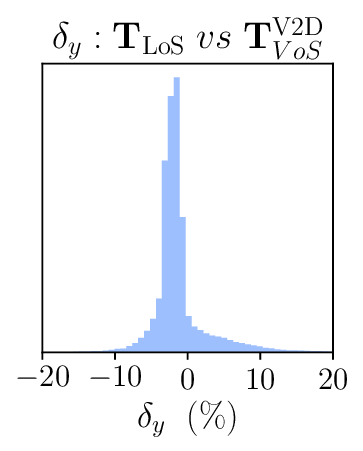}
        };
        \node[rotate=90, anchor=south, font=\small] at ([xshift=-0.25mm]img_d.west)
            {$\mathrm{frequency}$};
        \node[anchor=north, font=\small] at (img_d.south)
            {(d)};

        \node[anchor=north west, inner sep=0] (img_e) at (0.35\textwidth,-7.52) {
            \includegraphics[width=0.30\textwidth]{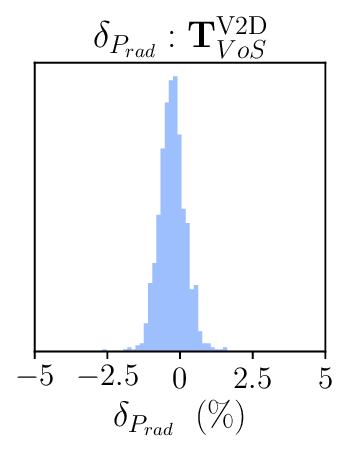}
        };
        \node[anchor=north, font=\small] at (img_e.south)
            {(e)};

        \node[anchor=north west, inner sep=0] (img_f) at (0.66\textwidth,-7.56) {
            \includegraphics[width=0.30\textwidth]{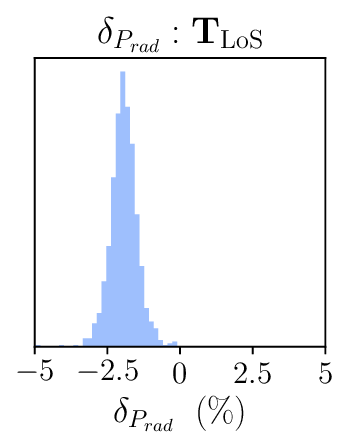}
        };
        \node[anchor=north, font=\small] at (img_f.south)
            {(f)};
    \end{tikzpicture}
    \captionsetup{justification=raggedright, singlelinecheck=false}
        \caption{Impact of geometry matrix accuracy on tomographic reconstruction: phantom study for TCV's SXR diagnostic. In Fig.\ref{fig:sxr_los_tmat}a, the line-of-sight configu- ration of the diagnostic. In Figs.\ref{fig:sxr_los_tmat}b and \ref{fig:sxr_los_tmat}c, visualization of the LoS ($\mathbf{T}_{LoS}$) and VoS ($\mathbf{T}_{VoS}^{V2D}$) geometry matrices, for a single detector from the top camera and for all detectors, respectively. In Fig.\ref{fig:sxr_los_tmat}d, relative error $\delta_y$ committed by estimating the measurements $\mathbf{y}$ using $\mathbf{T}_{LoS}$ instead of $\mathbf{T}_{VoS}^{V2D}$. The correct measurements $\mathbf{y}_{VoS}$ are then used to compute $\mathbf{x}_{_{MAP}}^{VoS}$ (using $\mathbf{T}_{VoS}^{V2D}$) and $\mathbf{x}_{_{MAP}}^{LoS}$ (using $\mathbf{T}_{LoS}$): in Figs.\ref{fig:sxr_los_tmat}e and \ref{fig:sxr_los_tmat}f, the relative errors $\delta_{P_{rad}}$ for the tomographically estimated SXR radiated powers $P_{rad}(\mathbf{x}_{_{MAP}}^{VoS})$ and $P_{rad}(\mathbf{x}_{_{MAP}}^{LoS})$, respectively.}
    \label{fig:sxr_los_tmat}
\end{figure}

We conduct the SXR study on the $10^3$ SXR emissivity phantoms $\mathcal{S}$ described in \cite{hamm_tomography_2025}. For each phantom, we follow the procedure discussed earlier in this section. In Figs.\ref{fig:sxr_los_tmat}a-\ref{fig:sxr_los_tmat}c we show the SXR diagnostic and the two geometry matrices $\mathbf{T}_{LoS}, \mathbf{T}_{VoS}^{V2D}$.
Visual comparison clearly reveals the smoother appearance of $\mathbf{T}_{VoS}^{V2D}$: it captures the increasing beam width and decreasing geometry values as the distance from the detector increases, whereas $\mathbf{T}_{LoS}$ does not. In Fig.\ref{fig:sxr_los_tmat}d, we show the distribution of the relative error $\delta_y=(y_{LoS}-y_{VoS})/y_{VoS}$. Its average value over all phantoms and detectors is $\widetilde{\delta}_y=-0.71\%\pm6.60\%$.
In Fig.\ref{fig:sxr_los_tmat}e we plot the distribution of the relative error $\delta_{P_{rad}}^{VoS}=(P_{rad}(\mathbf{x}_{_{MAP}}^{VoS})-P_{rad}^{true})/P_{rad}^{true}$, tomographically estimated with the reconstruction model using the correct matrix $\mathbf{T}_{VoS}^{V2D}$. The estimate $P_{rad}(\mathbf{x}_{_{MAP}}^{VoS})$ of the SXR radiated power is very accurate and precise, with almost no bias: the average relative error is $\widetilde{\delta}_{P_{rad}}^{VoS}=-0.27\%\pm0.46\%$,. In Fig.\ref{fig:sxr_los_tmat}f, we show the distribution of $\delta_{P_{rad}}^{LoS}=(P_{rad}(\mathbf{x}_{_{MAP}}^{LoS})-P_{rad}^{true})/P_{rad}^{true}$, where $\mathbf{x}_{_{MAP}}^{LoS}$ is computed using the mismatched model $\mathbf{T}_{LoS}$. The estimate $P_{rad}(\mathbf{x}_{_{MAP}}^{LoS})$ exhibits a small amount of negative bias, with $\widetilde{\delta}_{P_{rad}}^{LoS}=-1.93\%\pm0.47\%$. While the relative errors on the tomographic measurements can reach up to $\pm 10\%$ in the most extreme cases, $\delta_{P_{rad}}^{LoS}$ is typically much smaller. Finally, we find that the MSE ${E}_{MAP}^{LoS}$ is, on average, $41.37\%\pm23.70\%$ higher than ${E}_{MAP}^{VoS}$. The results show that using the LoS model for reconstruction leads to an overall poorer performance, in terms of both MSE and error on the SXR radiated power.

\subsection{Bolometry reconstruction}\label{bolo_effect}
For the bolometry study, we rely on a dataset $\mathcal{X}$ composed of $10^3$ physically realistic phantoms, whose generation is described in \cite[Appendix A]{hamm2026}.
The synthetic emissivity profiles are generated by combining SOLPS \cite{wiesen_new_2015} plasma boundary simulations with other physically realistic radiation features, such as core radiation and X-point radiator features.
We repeat the study from Section \ref{sxr_effect} on these $10^3$ phantoms.
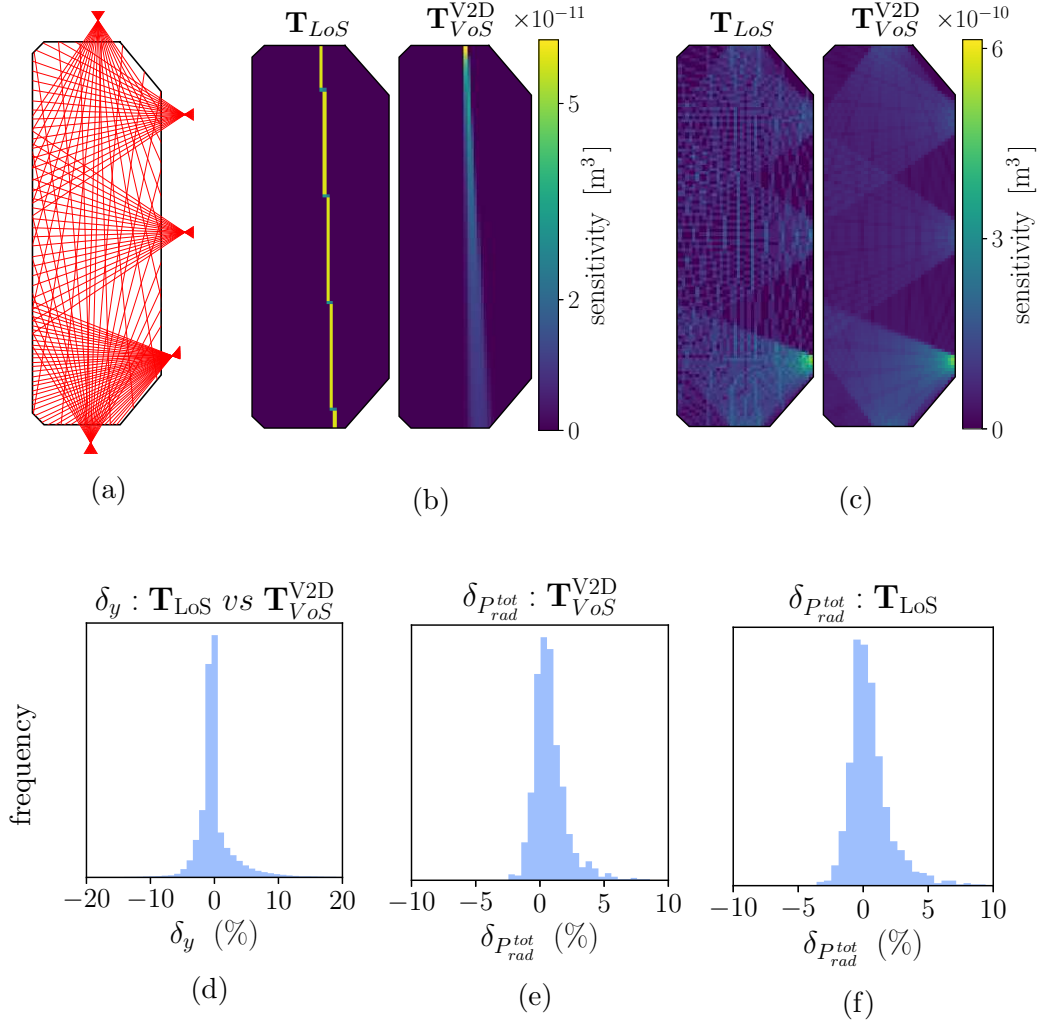
\begin{figure}[htbp]
    \centering
    \vspace{1.25cm}
    \begin{tikzpicture}
        \node[anchor=north west, inner sep=0] (img_a) at (0,0) {
            \includegraphics[width=0.188\textwidth]{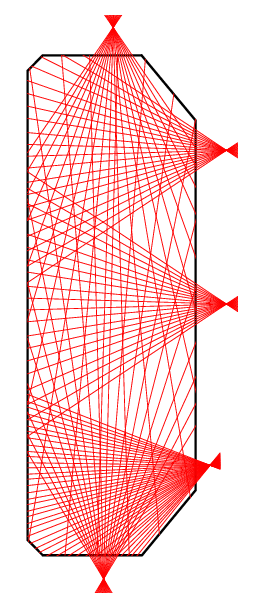}
        };
        \node[anchor=north, font=\small] at (img_a.south) {(a)};

        \node[anchor=north west, inner sep=0] (img_b) at (0.22\textwidth,0) {
            \includegraphics[width=0.37\textwidth]{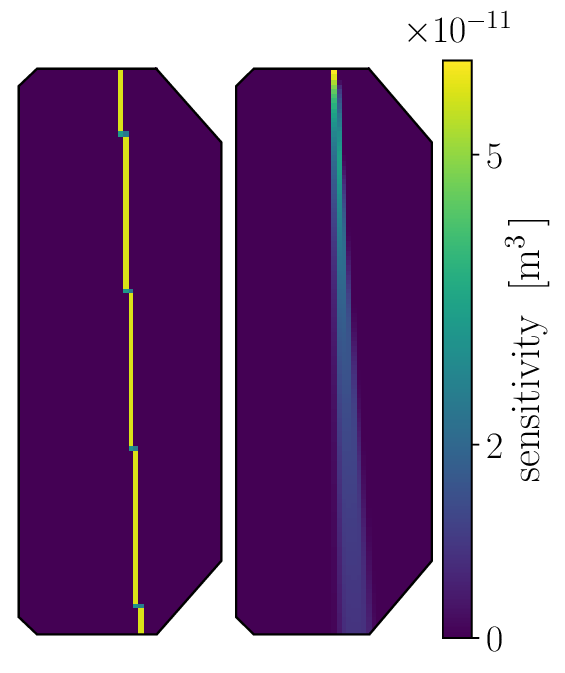}
        };
        \node[anchor=north, font=\small] at ([yshift=-2.9mm]img_b.south) {(b)};
        
        \node[anchor=south, font=\small] at ([xshift=-15mm, yshift=-6.5mm]img_b.north){$\mathbf{T}_{LoS}$};
        \node[anchor=south, font=\small] at ([xshift=4mm, yshift=-6.5mm]img_b.north){$\mathbf{T}_{VoS}^{\mathrm{V2D}}$};

        \node[anchor=north west, inner sep=0] (img_c) at (0.63\textwidth,0) {
            \includegraphics[width=0.37\textwidth]{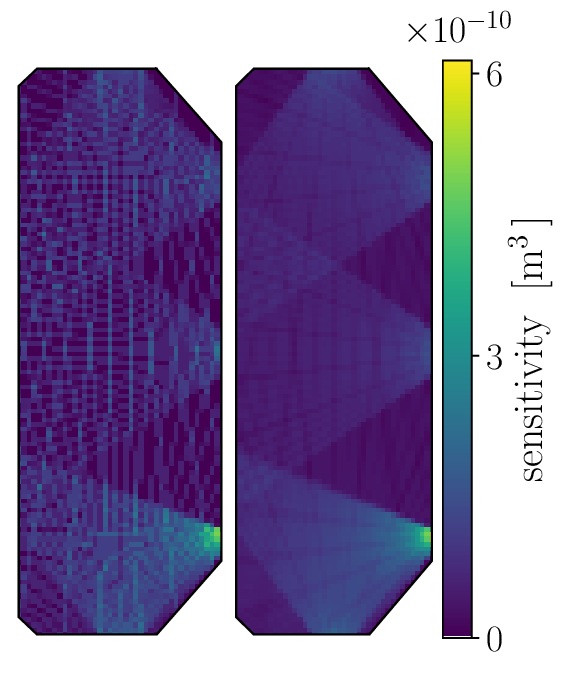}
        };
        \node[anchor=north, font=\small] at ([yshift=-2.9mm]img_c.south) {(c)};
        \node[anchor=south, font=\small] at ([xshift=-15mm, yshift=-6.5mm]img_c.north){$\mathbf{T}_{LoS}$};
        \node[anchor=south, font=\small] at ([xshift=4mm, yshift=-6.5mm]img_c.north){$\mathbf{T}_{VoS}^{\mathrm{V2D}}$};

        \node[anchor=north west, inner sep=0] (img_d) at (0.5,-7.5) {
            \includegraphics[width=0.31\textwidth]{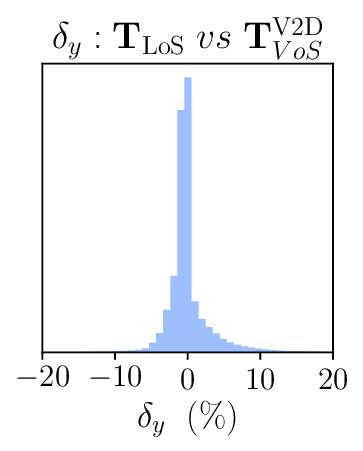}
        };
        \node[rotate=90, anchor=south, font=\small] at ([xshift=-0.25mm]img_d.west)
            {$\mathrm{frequency}$};
        \node[anchor=north, font=\small] at (img_d.south)
            {(d)};

        \node[anchor=north west, inner sep=0] (img_e) at (0.35\textwidth,-7.46) {
            \includegraphics[width=0.311\textwidth]{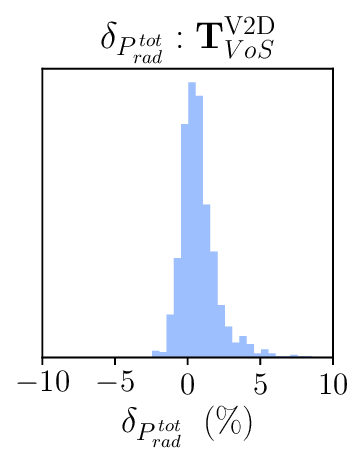}
        };
        \node[anchor=north, font=\small] at (img_e.south)
            {(e)};

        \node[anchor=north west, inner sep=0] (img_f) at (0.66\textwidth,-7.56) {
            \includegraphics[width=0.315\textwidth]{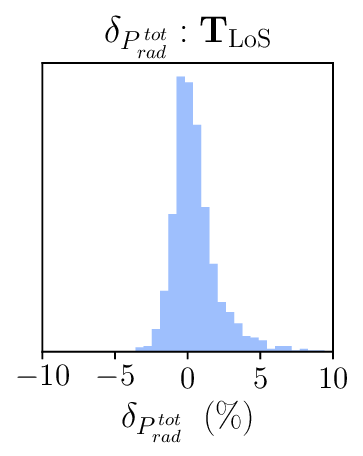}
        };
        \node[anchor=north, font=\small] at (img_f.south)
            {(f)};
    \end{tikzpicture}
    \captionsetup{justification=raggedright, singlelinecheck=false}
        \caption{Impact of geometry matrix accuracy on tomographic reconstruction: phantom study for TCV's bolometry diagnostic, shown in Fig.\ref{fig:bolo_los_tmat}a. In Figs.\ref{fig:bolo_los_tmat}b and \ref{fig:bolo_los_tmat}c, visualization of the LoS ($\mathbf{T}_{LoS}$) and VoS ($\mathbf{T}_{VoS}^{V2D}$) geometry matrices, for a single detector from the top camera and for all detectors, respectively. In Fig.\ref{fig:bolo_los_tmat}d, relative error $\delta_y$ committed by estimating the measurements $\mathbf{y}$ using $\mathbf{T}_{LoS}$ instead of $\mathbf{T}_{VoS}^{V2D}$. The correct measurements $\mathbf{y}_{VoS}$ are then used to compute $\mathbf{x}_{_{MAP}}^{VoS}$ (using $\mathbf{T}_{VoS}^{V2D}$) and $\mathbf{x}_{_{MAP}}^{LoS}$ (using $\mathbf{T}_{LoS}$): in Figs.\ref{fig:bolo_los_tmat}e and \ref{fig:bolo_los_tmat}f, the relative errors $\delta_{P_{rad}^{tot}}$ for the tomographically estimated total radiated powers $P_{rad}^{tot}(\mathbf{x}_{_{MAP}}^{VoS})$ and $P_{rad}^{tot}(\mathbf{x}_{_{MAP}}^{LoS})$, respectively.}
    \label{fig:bolo_los_tmat}
\end{figure}
In Figs.\ref{fig:bolo_los_tmat}a-\ref{fig:bolo_los_tmat}c we show the bolometry diagnostic and the geometry matrices $\mathbf{T}_{LoS}$ , $\mathbf{T}_{VoS}^{V2D}$. As for the SXR diagnostic, the bolometry VoS matrix is smoother and more physically realistic than the LoS one. In Fig.\ref{fig:bolo_los_tmat}d, we show the relative error $\delta_y=(y_{LoS}-y_{VoS})/y_{VoS}$ on the tomographic measurements. On average, $\widetilde{\delta}_y=0.15\%\pm3.89\%$. Compared to the SXR case, the error introduced by approximating the true VoS measurements with the LoS model is thus smaller. In Figs.\ref{fig:bolo_los_tmat}e and \ref{fig:bolo_los_tmat}f, we plot the distribution of the relative errors on the total radiated power $\delta_{P_{rad}^{tot}}^{VoS}$ and $\delta_{P_{rad}^{tot}}^{LoS}$, respectively. The two histograms are quite similar, suggesting that using the simpler LoS model for reconstruction does not strongly affect the estimation.
In Table \ref{table:prads_rel_errs_vos_los}, we report the average values and standard deviations of the relative errors on the tomographically estimated total, core, divertor and main chamber radiated powers. The divertor and main chamber, for a lower single null magnetic configuration, are defined as the regions below and above the X-point, respectively. The results shown in Table \ref{table:prads_rel_errs_vos_los} confirm the observation drawn from the histograms. The $P_{rad}$ estimates from the two models are similarly accurate, with the VoS ones being slightly more precise (smaller standard deviation). The two MAPs $\mathbf{x}_{_{MAP}}^{VoS}$ and $\mathbf{x}_{_{MAP}}^{LoS}$ lead to very similar estimates of the power radiated from the considered regions of interest. The simpler LoS model still leads to a larger MSE: $E_{_{MAP}}^{LoS}$ is, on average, $9\%\pm6\%$ higher than $E_{_{MAP}}^{VoS}$. In conclusion, the results show that using the correct geometry matrix for reconstruction leads to overall better performance, while also showing that using the mismatched geometry matrix does not significantly affect the estimation of the power radiated from large plasma regions. This shows that, upon proper regularization, the impact of using the simpler LoS geometry matrix is more limited than what the strong visual differences observable in Figs.\ref{fig:bolo_los_tmat}b and \ref{fig:bolo_los_tmat}c would suggest. The better performance in terms of MSE and $P_{rad}$ precision, however, still suggests to rely on the physically accurate VoS geometry matrix for reconstruction. 
\vspace{0.75cm}
\renewcommand{\arraystretch}{1.9}
\begin{table}[t]
\scalebox{0.95}{
  \begin{tabularx}{1.04\textwidth}{c|c|c|c|c|}
    \cline{2-5}
    & $\widetilde{\delta}_{P_{\mathrm{rad}}^{\mathrm{\,tot}}}$
    & $\widetilde{\delta}_{P_{\mathrm{rad}}^{\mathrm{\,core}}}$
    & $\widetilde{\delta}_{P_{\mathrm{rad}}^{\mathrm{\,div}}}$
    & $\widetilde{\delta}_{P_{\mathrm{rad}}^{\mathrm{\,main}}}$ \\ \hline
    VoS & $0.78\%\pm1.36\%$ & $0.21\%\pm4.63\%$ & $0.64\%\pm4.55\%$ & $1.07\%\pm3.09\%$ \\ \hline
    LoS & $0.53\%\pm1.67\%$ & $-0.25\%\pm5.75\%$ & $-0.67\%\pm4.64\%$ & $1.05\%\pm3.67\%$ \\ \hline
  \end{tabularx}}
  \captionsetup{justification=raggedright, singlelinecheck=false}
  \caption{Relative error on the total, core, divertor and main chamber radiated power when using the VoS or LoS geometry matrix in the reconstruction model.}
  \label{table:prads_rel_errs_vos_los}
  \end{table}

\section{Ray-tracing: description and comparison with voxel-to-detector approaches}\label{vos_vs_ray}
In this section, we first describe the ray-tracing approach to the computation of the geometry matrix. Then, we compare a Cherab-based ray-tracing implementation of RADCAM's bolometer diagnostic to the voxel-to-detector approach presented in Section \ref{v2d_approach}.
\subsection{Ray-tracing approach: theoretical foundations}
We closely follow \cite{carr_description_2018}: the reader is referred to it for a more complete discussion. Rather than estimating the measurements from discretized versions of Eq.\eqref{sensitivity_function}, ray-tracing approaches take their premises from the observation that the power measured by the detector can also be written as
\begin{equation}\label{raytracing_eq}
    y = \int_{A_{detector}}\int_{\Omega} L(\mathbf{s}, \omega)\,\cos\theta(\omega)\;\mathrm{d}\omega\;\mathrm{d}\mathbf{s}\;,
\end{equation}
where $L(\mathbf{s},\,\omega)$ is the incident radiance reaching the detector surface at the location $\mathbf{s}$ with incident angle $\omega$.
Here $\omega$ denotes a direction on the unit sphere and $\mathrm{d}\omega$ the associated solid-angle element, with $\Omega$ the solid angle spanning the incident directions; $\omega$ is commonly referred to as the incident angle, since $L(\cdot,\,\omega)$ is the radiance reaching the detector from that direction.
The $\cos\theta(\omega)$ factor in Eq.\eqref{raytracing_eq} accounts for the dependence of the measured power on the incident angle, which depends on the effective observing area. This area decreases as the incident angle $\omega$ deviates from the normal to the detector surface: $\cos\theta=1$ for a ray perpendicular to the surface, while $\cos\theta=0$ for a ray parallel to the surface. The radiance is expressed in units of $\mathrm{W/(m^2\,sr)}$. Eq.\eqref{raytracing_eq}, intuitively, states that the power reaching the detector can be computed by integrating the incident radiance over the surface of the detector and over all possible incident angles.\\
To estimate Eq.\eqref{raytracing_eq}, ray-tracing approaches implement a Monte Carlo (MC) estimator with importance sampling \cite{kroese_handbook_2011}. This class of MC methods can be used to estimate the integral $I =\int_\mathcal{V}\,f(\mathbf{x})\,\mathrm{d}\mathbf{x}$ of a function $f:\mathcal{V}\to\mathbb{R}$, with $\mathcal{V}\subseteq\mathbb{R}^{d}$, $d\geq1$. To estimate $I$, the function $f$ is evaluated at $N_{MC}$ points $\{\mathbf{x}_i\}_{i=1}^{N_{MC}}$. These points are sampled from a probability distribution $p_{MC}(\mathbf{x})$, with $p_{MC}:\mathcal{V}\to\mathbb{R}$, that is chosen to reduce the variance of the MC estimator; this effective and popular strategy is called importance sampling. The MC estimator $I_{MC}$ is obtained by averaging the integrand values $\{f(\mathbf{x}_i)\}_{i=1}^{N_{MC}}$ while accounting for their relative sampling frequency, which can be achieved by weighing each sample according to $\{p_{MC}(\mathbf{x}_i)\}_{i=1}^{N_{MC}}$. The resulting estimator reads
\begin{equation}\label{mc_integrator}
    I_{MC} = \frac{1}{N_{MC}}\sum_{i=1}^{N_{MC}}\frac{f(\mathbf{x}_i)}{p_{MC}(\mathbf{x}_i)}\;.
\end{equation}
The error $\vert I-I_{MC}\vert$ can be shown to decrease as $1/\sqrt{N_{MC}}\;$: it can thus be controlled by increasing the number of samples.\\
The MC estimator $y_{MC}$ for Eq.\eqref{raytracing_eq} is thus obtained by evaluating the integrand function $L(\mathbf{s}, \omega)\,\cos\theta(\omega)$ for $N_{MC}$ rays $\{(\mathbf{s}_i,\,\omega_i)\}_{i=1}^{N_{MC}}$, where each ray is parametrized in terms of its incident location $\mathbf{s}_i$ and incidence angle $\omega_i$. Importance sampling is applied by sampling $\mathbf{s}_i$ and $\omega_i$ from uniform probability distributions over the detector surface and over a solid angle $\Omega_{frac}$, respectively; $\Omega_{frac}$ is chosen by defining a suitable minimally viewing cone, to minimize the number of rays that do not intersect the pinhole. The corresponding importance sampling probability density function is $p_{MC}(\mathbf{s},\omega)=1/(A_{detector}\,\Omega_{frac})$, $\forall \;(\mathbf{s},\omega)\in A_{detector} \times \Omega_{frac}$. Ultimately, $y_{MC}$ is given by
\begin{equation}\label{y_mc_eq}
    y_{MC} = \frac{\Omega_{frac}A_{detector}}{N_r}\sum_{i=1}^{N_{MC}}L(\mathbf{s}_i, \omega_i)\,\cos\theta(\omega_i)\;,
\end{equation}
where the incident radiance $L(\mathbf{s}_i, \omega_i)$ is the line integral of the emissivity function $\varepsilon_{\Omega}$ (with $\varepsilon_{\Omega}$ in units of $\mathrm{W/(m^3\,sr)}$) along the direction defined by the ray $(\mathbf{s}_i, \omega_i)$. From Eq.\eqref{y_mc_eq}, we see that $y_{MC}$ is a weighted average of line-integrated emissivity values over the sampled rays. This key observation, as discussed in \cite{carr_description_2018}, allows the definition of a geometry matrix $\mathbf{T}_{VoS}^{\mathrm{RayT}, N_{MC}}\in\mathbb{R}^{M\times N_zN_r}$. First, one defines a grid of $N_zN_r$ toroidally symmetric voxels, built as those defined in Section \ref{v2d_approach} but obtained through a full toroidal $2\pi$ rotation, so that $\phi = 2\pi$. Then, for each detector $m=1,\ldots,M$, one samples $N_{MC}$ rays starting from it and traces their paths through the vessel, computing for each ray $i=1,\ldots,N_{MC}$ the length $s_{i,m,j}$ of its intersection with each voxel $j=1,\ldots,N_zN_r$. Finally, the elements of $\mathbf{T}_{VoS}^{\mathrm{RayT}, N_{MC}}$ are computed as
\begin{equation}\label{raytracing_gmat_elements}
    \big(\mathbf{T}_{VoS}^{\mathrm{RayT}, N_{MC}}\big)_{mj} = \frac{\Omega_{frac}A_{detector}}{4\pi\,N_r}\sum_{i=1}^{N_{MC}}s_{i,m,j}\,\cos\theta(\omega_i)\;,
\end{equation}
where the factor $4\pi$ is introduced to obtain a geometry matrix in units of $\mathrm{m^3}$. Ray-tracing can thus be understood as a detector-to-voxel approach, following a reverse logic compared to the voxel-to-detector one described in Section \ref{v2d_approach}.\\
Intuitively, in the ideal limit of infinitely small pinhole and detector, from Eq.\eqref{raytracing_gmat_elements} we recover the LoS approximation from Eq.\eqref{los_vos_equivalence}. In this ideal case, all the $N_{MC}$ sampled rays coincide with the beam centerline (the sightline), so that $\cos\theta(\omega_i)=\cos\beta$ for all rays, with $\beta$ the angle between the sightline and detector normal. Moreover, $s_{i,m,j}=s_{m,j}$ for all rays, therefore $1/N_r\sum_is_{i,m,j}=s_{m,j}$ because all rays follow the same path. Eq.\eqref{raytracing_gmat_elements} thus simplifies to $\big(\mathbf{T}_{VoS, \;\mathrm{ideal}}^{\mathrm{RayT}, N_{MC}}\big)_{mj} = \frac{\Omega_{frac}A_{detector}\cos\beta}{4\pi}s_{m,j}$. In the ideal case, the solid angle $\Omega_{frac}$ corresponding to the minimally viewing cone is simply given by $(A_{pinhole}\cos\alpha)/L^2$, with $\alpha$ the angle between sightline and pinhole normal and $L$ the detector-pinhole distance. The matrix elements are thus given by $\big(\mathbf{T}_{VoS, \;\mathrm{ideal}}^{\mathrm{RayT}, N_{MC}}\big)_{mj} = \widehat{\mathcal{E}}_{LoS}\;s_{m,j}$, with $\widehat{\mathcal{E}}_{LoS}$ the etendue approximation from Eq.\eqref{etendue_los}.
\begin{figure}[t!]
    \centering
    \begin{tikzpicture}
        \node[anchor=north west, inner sep=0] (img_a) at (0,0) {
            \includegraphics[width=0.188\textwidth]{bolo_LoS_configuration.eps}
        };
        \node[anchor=north, font=\small] at (img_a.south) {(a)};

        \node[anchor=north west, inner sep=0] (img_b) at (0.22\textwidth,0) {
            \includegraphics[width=0.37\textwidth]{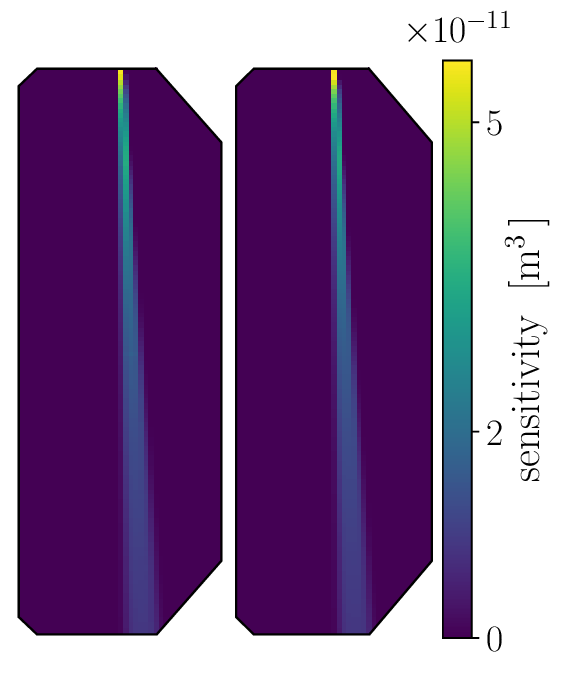}
        };
        \node[anchor=north, font=\small] at (img_b.south) {(b)};
        \node[anchor=south, font=\small] at ([xshift=-15mm, yshift=-6.5mm]img_b.north){$\mathbf{T}_{VoS}^{\mathrm{RayT},\,10^5}$};
        \node[anchor=south, font=\small] at ([xshift=4mm, yshift=-6.5mm]img_b.north){$\mathbf{T}_{VoS}^{\mathrm{V2D},7}$};

        \node[anchor=north west, inner sep=0] (img_c) at (0.63\textwidth,0) {
            \includegraphics[width=0.37\textwidth]{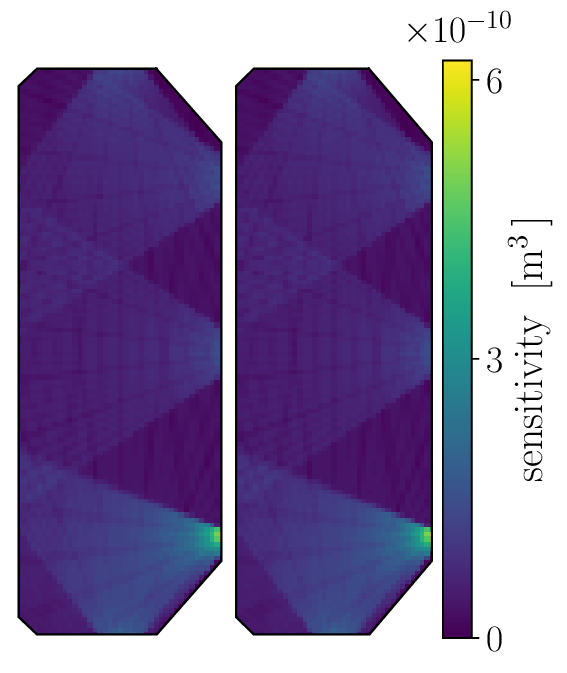}
        };
        \node[anchor=north, font=\small] at (img_c.south) {(c)};
        \node[anchor=south, font=\small] at ([xshift=-15mm, yshift=-6.5mm]img_c.north){$\mathbf{T}_{VoS}^{\mathrm{RayT},\,10^5}$};
        \node[anchor=south, font=\small] at ([xshift=4mm, yshift=-6.5mm]img_c.north){$\mathbf{T}_{VoS}^{\mathrm{V2D},7}$};
    \end{tikzpicture}
      \captionsetup{justification=raggedright, singlelinecheck=false}
    \caption{Ray-tracing vs voxel-to-detector approach. In Fig.\ref{fig:bolo_ray_vs_vos_tmat}a, line-of-sight configuration of the bolometry diagnostic. In Fig.\ref{fig:bolo_ray_vs_vos_tmat}b and \ref{fig:bolo_ray_vs_vos_tmat}c, a side-by-side comparison of the ray-tracing ($\mathbf{T}_{VoS}^{\mathrm{RayT},\,10^5}$) and voxel-to-detector ($\mathbf{T}_{VoS}^{\mathrm{V2D}, 7}$) geom- etry matrices, for a single detector from the top camera and for all detectors, respectively.}
    \label{fig:bolo_ray_vs_vos_tmat}
\end{figure}
\subsection{Comparison with voxel-to-detector approach}
\begin{figure}[t!]
\centering
\begin{tikzpicture}
  \def\pw{0.15\textwidth} 
  \def\gap{0.035\textwidth}
  \def\cbw{0.025\textwidth}
  \def\cbgap{0.9\textwidth}
  \def\rowgap{-0.09\textwidth}
  \def\pwb{0.42\textwidth}
  \def\gapb{0.06\textwidth}

  \foreach \i in {0,1,2,3,4} {
    \node[inner sep=0pt, anchor=north west] (plot\i) at ({(\i)*(\pw+\gap)}, 0) {
      \includegraphics[width=\pw]{rel_diff_\i.eps}
    };
  }

\node[inner sep=0pt, anchor= center] (colorbar) at ([xshift=10.2mm, yshift=-1mm]plot4.east) {
            \includegraphics[height=0.4\textwidth]{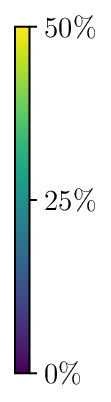}
        };

    \node[anchor=north] at
    ([xshift=1mm, yshift=0mm]plot2.south) {
    (a) relative pixel-wise difference: $\mathbf{T}_{VoS}^{\mathrm{RayT},\,N_{MC}}\;vs\;\mathbf{T}_{VoS}^{\mathrm{V2D},7}$
  };

  \node[anchor=center, font=\small] at ([xshift=0mm, yshift=0.mm]plot0.north){${N_{MC}}=10^1$};
\node[anchor=center, font=\small] at ([xshift=0mm, yshift=0mm]plot1.north){${N_{MC}}=10^2$};
  \node[anchor=center, font=\small] at ([xshift=0mm, yshift=0mm]plot2.north){${N_{MC}}=10^3$};
    \node[anchor=center, font=\small] at ([xshift=0mm, yshift=0mm]plot3.north){${N_{MC}}=10^4$};
      \node[anchor=center, font=\small] at ([xshift=0mm, yshift=0mm]plot4.north){${N_{MC}}=10^5$};

  \node[inner sep=0pt, anchor=north] (plotb0) at
    ([yshift=-12.75mm, xshift=18mm]plot0.south) {
    \includegraphics[width=\pwb]{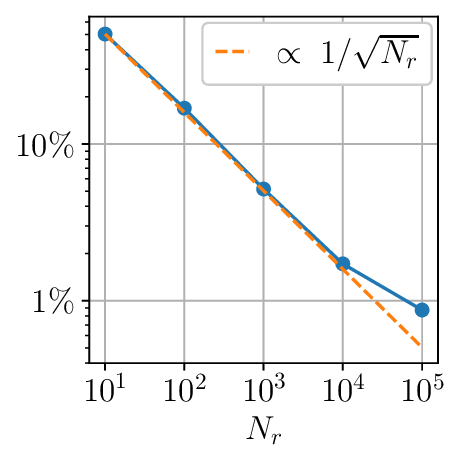}
  };

  \node[inner sep=0pt, anchor=north] (plotb1) at
    ([yshift=-12.75mm, xshift=10mm]plot3.south) {
    \includegraphics[width=\pwb]{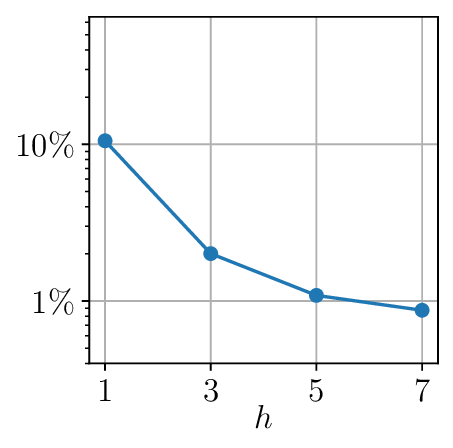}
  };
    \node[anchor=north,align=center] at ([yshift=2mm, xshift=2mm]plotb0.south){
    (b) relative norm difference:\\
    \vspace{-0.5cm}\\
$\mathbf{T}_{VoS}^{\mathrm{RayT},\,N_{MC}}\;vs\;\mathbf{T}_{VoS}^{\mathrm{V2D},7}$
  };
  \node[anchor=north, align=center] at ([yshift=2mm]plotb1.south){
    (c) relative norm difference:\\
    \vspace{-0.5cm}\\
$\;\mathbf{T}_{VoS}^{\mathrm{V2D},\,h}\;vs\;\mathbf{T}_{VoS}^{\mathrm{RayT},\,10^5}$
  };
\end{tikzpicture}
\captionsetup{justification=raggedright, singlelinecheck=false}
\caption{Ray-tracing vs voxel-to-detector approach: convergence properties as a function of number of sampled rays $N_{MC}$ and grid refinement factor $h$. In Fig.\ref{fig:bolo_ray_vs_vos_tmat_rel_err}a, relative pixel-wise difference between $\mathbf{T}_{VoS}^{\mathrm{RayT},\,N_{MC}}$ and $\mathbf{T}_{VoS}^{\mathrm{V2D}, 7}$. In Fig.\ref{fig:bolo_ray_vs_vos_tmat_rel_err}b, relative norm difference between the same matrices $\mathbf{T}_{VoS}^{\mathrm{RayT},\,N_{MC}}$ and $\mathbf{T}_{VoS}^{\mathrm{V2D}, 7}$. In Fig.\ref{fig:bolo_ray_vs_vos_tmat_rel_err}c, relative norm difference between $\mathbf{T}_{VoS}^{\mathrm{V2D}, h}$ and $\mathbf{T}_{VoS}^{\mathrm{RayT},\,10^5}$.
}
    \label{fig:bolo_ray_vs_vos_tmat_rel_err}
\end{figure}

Let us consider the bolometer diagnostic discussed in Section \ref{bolo_effect}. We compare the synthetic diagnostics obtained with the ray-tracing and voxel-to-detector models to investigate whether the two approaches agree and to study their convergence properties. Relying on Cherab, we compute the ray-tracing geometry matrices $\mathbf{T}_{VoS}^{\mathrm{RayT},\,N_{MC}}$ for $N_{MC}=10^1,10^2,10^3,10^4,10^5$ rays. Moreover, we compute the voxel-to-detector matrices $\mathbf{T}_{VoS}^{\mathrm{V2D}, h}$ for increasing grid refinement factors $h=1,3,5,7$. In both cases, the matrices are defined on the same coarse $N_z\times N_r$ poloidal grid, allowing a direct pixel-to-pixel comparison.\\
In Fig.\ref{fig:bolo_ray_vs_vos_tmat}, we compare the most accurate matrices obtained with each approach ($N_{MC}=10^5$ for ray-tracing and $h=7$ for V2D). While some small differences are visible in Fig.\ref{fig:bolo_ray_vs_vos_tmat}b, particularly near the pinhole, the two matrices are in very good agreement. By summing together the information for all detectors as in Fig.\ref{fig:bolo_ray_vs_vos_tmat}c, in particular, the two matrices become essentially indistinguishable.\\
To quantify the difference between the matrices as a function of their parameters $N_{MC}$ and $h$, we introduce the following relative difference metrics. Given two arbitrary matrices $\mathbf{A}$ and $\mathbf{B}$, we define their relative pixel-wise difference as $\delta_{ij}=\vert A_{ij}-B_{ij}\vert/B_{ij}$ for all pixels $(i,j)$, and their relative (Frobenius) norm difference as $\;\delta_{L^2}=\sqrt{\sum_{i,j}(A_{ij}-B_{ij})^2}/\sqrt{\sum_{i,j}B_{ij}^2}\;$. To compare the ray-tracing and V2D approaches, we then consider the following metrics: the relative pixel-wise difference between $\mathbf{T}_{VoS}^{\mathrm{RayT},\,N_{MC}}$ and $\mathbf{T}_{VoS}^{\mathrm{V2D}, 7}$ as a function of $N_{MC}$, the relative norm difference between $\mathbf{T}_{VoS}^{\mathrm{RayT},\,N_{MC}}$ and $\mathbf{T}_{VoS}^{\mathrm{V2D}, 7}$ as a function of $N_{MC}$, and the relative norm difference between $\mathbf{T}_{VoS}^{\mathrm{V2D}, h}$ and $\mathbf{T}_{VoS}^{\mathrm{RayT},\,10^5}$ as a function of $h$.\\
In Fig.\ref{fig:bolo_ray_vs_vos_tmat_rel_err}a we show the decrease, as the number of rays $N_{MC}$ increases, of the relative pixel-wise difference between the geometry matrices $\mathbf{T}_{VoS}^{\mathrm{RayT},\,N_{MC}}$ and the most accurate V2D matrix $\mathbf{T}_{VoS}^{\mathrm{V2D}, 7}$. To be precise, the relative difference is computed on the row-summed matrices obtained by summing the sensitivity maps of all detectors (as in Fig.\ref{fig:bolo_ray_vs_vos_tmat}c, e.g.). The plots in Fig.\ref{fig:bolo_ray_vs_vos_tmat_rel_err}a clearly show that the ray-tracing matrices become increasingly closer to $\mathbf{T}_{VoS}^{\mathrm{V2D}, 7}$ as $N_{MC}$ increases. For low $N_{MC}$ values, the relative difference is large: the number of sampled rays is not sufficient, leading to a poor MC estimator in Eq.\eqref{y_mc_eq} and thus to an inaccurate matrix in Eq.\eqref{raytracing_gmat_elements}. The plot corresponding to the largest considered value $N_{MC}=10^5$ corresponds to the relative pixel-wise difference of the matrices shown in Fig.\ref{fig:bolo_ray_vs_vos_tmat}c. The low relative difference for $N_{MC}=10^5$ confirms that the two matrices are very close.\\
$\!\!$In Fig.\ref{fig:bolo_ray_vs_vos_tmat_rel_err}b we show the decay, as $N_{MC}$ increases, of the relative norm difference between $\mathbf{T}_{VoS}^{\mathrm{RayT},\,N_{MC}}$ and $\mathbf{T}_{VoS}^{\mathrm{V2D}, 7}$. These are the same matrices considered in Fig.\ref{fig:bolo_ray_vs_vos_tmat_rel_err}a. The plot shows that, until $N_{MC}\approx10^4$, the decay rate of the relative norm difference is almost perfectly proportional to $1/\sqrt{N_{MC}}$, which is the expected MC decay rate. This suggests that the MC estimator is behaving correctly and converging to its limit value at the expected rate. In correspondence of $N_{MC}\approx10^4$, we observe a transition: the decay rate becomes slower than $1/\sqrt{N_{MC}}$ and the curve starts flattening. The curve suggests that, for low $N_{MC}$ values, the dominant error source is the inaccuracy due to insufficient MC sampling. If $\mathbf{T}_{VoS}^{\mathrm{V2D}, 7}$ corresponded to the true geometry matrix and if the ray-tracing matrices $\mathbf{T}_{VoS}^{\mathrm{RayT},\,N_{MC}}$ were converging to it, the decay rate would keep following $1/\sqrt{N_{MC}}$ also for large $N_{MC}$ values. The flattening of the curve suggests that, for large $N_{MC}$ values, the difference between the two approaches is mainly driven by systematic differences between the two approaches: $\mathbf{T}_{VoS}^{\mathrm{RayT},\,N_{MC}}$ is not converging to $\mathbf{T}_{VoS}^{\mathrm{V2D}, 7}$. This behavior could be due to the discretization error inherent to the V2D approach, which considers the solid angle to be constant over the volume of each voxel. Alternatively, it could be due to a small error in the implementation of the ray-tracing approach. Either way, the plot confirms once more that $\mathbf{T}_{VoS}^{\mathrm{RayT},\,10^5}$ and $\mathbf{T}_{VoS}^{\mathrm{V2D}, 7}$ are very similar, with relative norm difference smaller than $1\%$. Since the curve is still decreasing, the difference would settle around even smaller values if one were to compute geometry matrices for $N_{MC}>10^5$, indicating that the two approaches lead to very similar results.\\
In Fig.\ref{fig:bolo_ray_vs_vos_tmat_rel_err}c, we plot the relative norm difference between $\mathbf{T}_{VoS}^{\mathrm{V2D}, h}$ and the most accurate ray-tracing matrix $\mathbf{T}_{VoS}^{\mathrm{RayT},\,10^5}$ as a function of $h$. The difference decreases as $h$ (and thus the accuracy of the V2D matrix) increases. The curve, while still decreasing at $h=7$, appears to be flattening. It would settle around a plateau value if matrices for $h>7$ were computed. Once again, this observation seems to confirm that some small systematic differences would still remain even for large $h$.\\
These residual differences seem to reflect subtle implementation differences between the two approaches. However, importantly, the study shows that for sufficiently high values of $N_{MC}$ and $h$, both resulting matrices are expected to be of comparable accuracy. The residual differences, moreover, are not expected to significantly affect the quality of the tomographic reconstructions obtained using either of the two matrices. Indeed, by studying the relative difference $\delta_y^h=(y_{VoS}^{\mathrm{V2D, h}}-y_{VoS}^{\mathrm{RayT}, 10^5})/y_{VoS}^{\mathrm{RayT}, 10^5}$ on the tomographic data computed over a population of $10^3$ realistic phantoms (as in Section \ref{bolo_effect}), we obtain $\delta_y^h\approx0.01\%\pm0.78\%$ for both $h=5$ and $h=7$. Therefore, the residual norm differences do not lead to systematic differences in the computed tomography data. The very low average and standard deviation values of $\delta_y$ allow to safely conclude that the two models are expected to be, for all practical tomography purposes, essentially equivalent. The study discussed in this section provides confidence in the accuracy of both approaches. Following two specular approaches (voxel-to-detector vs detector-to-voxel), we obtain very similar models. The presented analysis validates the use of either of the two models for tomographic reconstruction of TCV data.

\section{Summary and discussion}\label{vos_summary}
In this paper, we discussed several techniques for computing synthetic diagnostics for plasma tomography. These synthetic diagnostics, typically referred to as geometry matrices, are the linear forward model $\mathbf{T}$ mapping emissivity profiles to tomographic data in the formulation of tomography inverse problems. A careful study of the accuracy of the computed geometry matrices is necessary to ensure that the obtained results are not only qualitatively but also quantitatively correct, thus allowing quantitative imaging. In Section \ref{v2d_approach}, we introduced an intuitive voxel-to-detector approach allowing the computation of physically accurate VoS geometry matrices. In Section \ref{los_vs_vos}, we compared the proposed approach to the simple LoS approximation. Conducting a phantom-based study on the SXR and bolometry TCV systems, we showed that using the more accurate VoS model is recommended, as it is expected to lead to overall better tomographic reconstruction. However, importantly, the study indicates that using the simpler LoS model is not expected to cause any significant systematic error in the estimation of the power radiated from large plasma regions. This is particularly crucial for bolometry, where obtaining quantities such as the total, core, divertor and main chamber radiated power is often one of the primary goals of tomographic reconstruction. Finally, in Section \ref{vos_vs_ray}, we compared the proposed voxel-to-detector model to a Cherab-based ray-tracing approach. We showed that the geometry matrices obtained with the two approaches agree very well, validating their routine use for tomographic reconstruction of TCV's SXR and bolometry data.\\
Efforts to adopt a more accurate VoS geometry matrix had been ongoing at TCV in recent years. However, the adoption of such a model had so far been prevented by the difficulty of independently verifying any attempt at generating such a model. The proposed V2D approach provides a simple, interpretable and easily inspectable validation tool for geometry matrices computed with other approaches. This allows, e.g., the validation of the ray-tracing implementation, increasing the confidence in the accuracy of the obtained models and providing a useful safeguard to implementation errors. Comparative convergence studies, such as those presented in Section \ref{vos_vs_ray}, are also useful to select adequate values of parameters such as the grid refinement factor $h$ (for the V2D approach) and the number of rays $N_{MC}$ (for the ray-tracing approach). The results show that both approaches can be used to obtain physically accurate geometry matrices. Since ray-tracing allows the implementation of wall reflections \cite{carr_physically_2019} and the handling of accurate CAD-based representations of the vessel and baffles, this approach may be preferable for some user cases. Ray-tracing is also, in general, more computationally efficient. This is not, however, a crucial factor, since geometry matrix computation needs to be done only once for a given diagnostic and vessel configuration.\\
The studies presented in this work assess the impact of using a geometry matrix more accurate than the simple LoS one used so far at TCV. Whether using the LoS matrix was introducing substantial systematic errors in the estimated quantities is a question that had been lingering around TCV's tomographic reconstructions. Our studies show that, upon proper regularization, the impact of using the LoS model is smaller than what is suggested by visual comparison to the VoS model (see Figs.\ref{fig:sxr_los_tmat}b-\ref{fig:sxr_los_tmat}c and \ref{fig:bolo_los_tmat}b-\ref{fig:bolo_los_tmat}c). Reconstruction using smaller amounts or less powerful regularization, as discussed in \cite{carr_description_2018}, could be more significantly impacted by the use of the LoS model. However, given the severely under-determined nature of plasma tomography, strong regularization is either way needed to obtain optimal results. Therefore, we conclude that the LoS model leads to overall worse results, as indicated by the larger MSEs of its reconstructions, but without introducing significant systematic biases in the estimation.
Since the computed VoS models were shown to be physically accurate and to reduce the MSE of the reconstructed emissivity profile, using the V2D or ray-tracing geometry matrices remains definitely preferable.

\section*{Acknowledgements}
This work was supported in part by the Swiss National Science Foundation. This work has been carried out within the framework of the EUROfusion Consortium, via the Euratom Research and Training Programme (Grant Agreement No 101052200 — EUROfusion) and funded by the Swiss State Secretariat for Education, Research and Innovation (SERI). Views and opinions expressed are however those of the author(s) only and do not necessarily reflect those of the European Union, the European Commission, or SERI. Neither the European Union nor the European Commission nor SERI can be held responsible for them.

\newpage

\bibliographystyle{unsrt}
\typeout{}
\bibliography{synthetic_diag}

\end{document}